\documentclass[pra,twocolumn,floatfix,footinbib,bibnotes,longbibliography]{revtex4-1}

\usepackage{graphicx}
\usepackage{epsfig,tabularx,multirow,booktabs}
\usepackage{amsmath,bbm,amssymb,bm,times}

\usepackage[colorlinks,linkcolor=blue,citecolor=blue]{hyperref}

\newcommand{\eqn}[1]{
\begin{eqnarray}
	#1
\end{eqnarray}
}

\newcolumntype{C}{>{$}c<{$}}
\AtBeginDocument{
\heavyrulewidth=.08em
\lightrulewidth=.05em
\cmidrulewidth=.03em
\belowrulesep=.65ex
\belowbottomsep=0pt
\aboverulesep=.4ex
\abovetopsep=0pt
\cmidrulesep=\doublerulesep
\cmidrulekern=.5em
\defaultaddspace=.5em
}

\begin{document}
\title{Nonperturbative Waveguide Quantum Electrodynamics}
\author{Yuto Ashida}
\email{ashida@phys.s.u-tokyo.ac.jp}
\affiliation{Department of Physics, University of Tokyo, 7-3-1 Hongo, Bunkyo-ku, Tokyo 113-0033, Japan}
\affiliation{Institute for Physics of Intelligence, University of Tokyo, 7-3-1 Hongo, Tokyo 113-0033, Japan}
\author{Takeru Yokota}
\affiliation{Interdisciplinary Theoretical and Mathematical Sciences Program (iTHEMS), RIKEN, Wako, Saitama 351-0198, Japan}
\affiliation{Institute for Solid State Physics, The University of Tokyo, Kashiwa, Chiba 277-8581, Japan}
\author{Ata\c c $\dot{\mathrm{I}}$mamo$\breve{\mathrm{g}}$lu}
\affiliation{Institute of Quantum Electronics, ETH Zurich, CH-8093 Z{\"u}rich, Switzerland}
\author{Eugene Demler}
\affiliation{Department of Physics, Harvard University, Cambridge, MA 02138, USA}

\begin{abstract} 
Understanding physical properties of quantum emitters strongly interacting with quantized electromagnetic modes is one of the primary goals in the emergent field of waveguide quantum electrodynamics (QED). When the light-matter coupling strength is comparable to or even exceeds energies of elementary excitations, conventional approaches based on perturbative treatment of light-matter interactions, two-level description of matter excitations, and photon-number truncation are no longer sufficient. Here we study in and out of equilibrium properties of waveguide QED in such nonperturbative regimes on the basis of a comprehensive and rigorous theoretical approach using an asymptotic decoupling unitary transformation. We uncover several surprising features ranging from symmetry-protected many-body bound states in the continuum to strong renormalization of the effective mass and potential; the latter may explain recent experiments demonstrating cavity-induced changes in chemical reactivity as well as enhancements of ferromagnetism or superconductivity. To illustrate our general results with concrete examples, we use our formalism to study a model of coupled cavity arrays, which is relevant to experiments in superconducting qubits interacting with microwave resonators or atoms coupled to photonic crystals. We examine the relation between our results and delocalization-localization transition in the spin-boson model; notably, we point out that a reentrant transition can occur in the regimes where the coupling strength becomes the dominant energy scale. We also discuss applications of our results to other problems in different fields, including quantum optics, condensed matter physics, and quantum chemistry. 
\end{abstract}

\maketitle

\section{Introduction}
\subsection{Background}

Quantum states arising from strong coherent interaction between light and matter are not only interesting from the perspective of fundamental many-body physics, but also provide promising new platforms for quantum technologies.  Historically, analysis of light-matter systems focused on 
 the perturbative regime \cite{RI37,CCT89},  since interaction of atomic dipoles with vacuum electromagnetic fields is weak due to the smallness of the fine structure constant $\alpha\!=\!1/137$. Recent progress has led to experimental realizations of new systems in which electromagnetic field is modified to reach stronger light-matter coupling. In particular, (artificial) atoms coupled to one-dimensional continuum of photons at  microwave \cite{BLS09,AO10,HIC11,HIC12,vLA13,MJA14,MM19,KB20,PR18} or optical \cite{RD13,TJD13,AMS14,YR14,GA14,LP15} frequencies achieve  enhancement of light-matter interaction through strong spatial confinement of electromagnetic modes. This rapidly growing field of research has been dubbed waveguide quantum electrodynamics (QED).

There exist many conceptual similarities between the questions addressed in waveguide QED and the problems analyzed in the context of quantum dissipative systems \cite{LAJ87,SA83,GF85,WU12}; in the latter, bosonic modes represent phonons or other collective excitations of condensed matter systems. More recently, light-matter interaction has also been the subject of intense research in the fields of polaritonic chemistry \cite{HJA12,GJ15,ETW16,HF16,FJ172,FJ17,Hiura2018,Hiura2019,TA19,STH21} and nanostructured plasmonics \cite{BW03,JMP06,TMS13,BJJ19,MNS20}. 
In light of such broad relevance, models of quantum emitters interacting with a continuum of bosonic excitations have played a crucial role in quantum information science as well as in condensed matter physics and quantum chemistry. An outstanding challenge here is to uncover the novel physical phenomena in {\it nonperturbative} regimes, where strong interaction leads to the formation of quantum many-body states with large entanglement among emitters and bosonic excitations of the continuum.

Despite recent remarkable advances, our understanding of waveguide QED at strong couplings is far from complete. Due to virtual excitation of many photons, the problem becomes intrinsically nonperturbative and standard approximations of quantum optics fail in many crucial aspects. First and foremost, it is known that the usual rotating wave approximation becomes no longer valid \cite{CCT89} due to the processes that create or annihilate pairs of excitations.  Moreover, the inclusion of the diamagnetic $\hat{A}^2$ term and the multilevel structure of emitters becomes more essential at larger coupling strengths. Importantly, the latter indicates that  the comprehensive understanding of strong coupling physics cannot be achieved unless one goes beyond the standard two-level descriptions. Such  multilevel structure of quantum emitters is also of current technological importance. For instance, superconducting transmon qubits are rarely operated as perfect two-level systems \cite{BA20} and such multilevel structures are potentially useful for the purpose of performing certain quantum information operations \cite{Rosenblum266,ESS20}.  

\begin{figure*}[t]
\includegraphics[width=160mm]{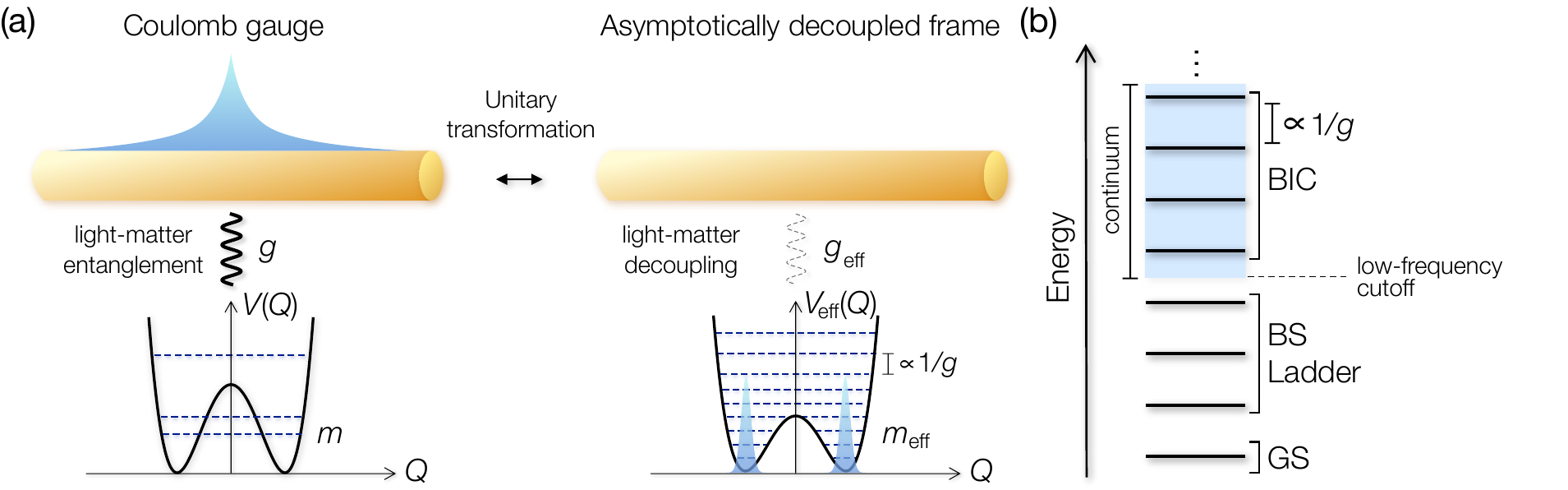} 
\caption{\label{fig_schem}
(a) Schematic illustration of the analysis. (Left) In the original Coulomb gauge, a single or multiple emitters interact with common electromagnetic continuum in arbitrary geometry via light-matter coupling $g$. A quantum emitter is modeled by a charged quantum particle of mass $m$ and position $Q$ that is trapped in a potential $V$. The potential is typically assumed to be a double-well potential as appropriate for an effectively two-level emitter though our theory is equally applicable to a generic  potential profile.  (Right) We use the newly introduced unitary transformation to asymptotically decouple emitter and photon degrees of freedom in the strong-coupling limit. After the transformation, emitters and photons interact with each other via vanishingly weak effective coupling that scales as $g_{\rm eff}\propto g^{-1/2}$ at large $g$. In contrast, the renormalized mass is enhanced as $m_{\rm eff}\!\propto\!g^2$, leading to the tight localization of the emitter at the potential minima as well as the $1/g$ energy spacing. The potential is renormalized to $V_{\rm eff}$ with lower potential barrier due to the dressing by the vacuum electromagnetic fluctuations. Note that, when going back to the Coulomb gauge, $Q$ in the asymptotically decoupled frame contains both matter and light contributions. (b) Formation of a ladder of the bound states (BS) and the bound states in the continuum (BIC) on top of the ground state (GS) in the nonperturbative regimes. The energy spacing and excitation energies  decrease as $\propto\! g^{-1}$ and thus, these states become increasingly degenerate at strong couplings. We note that there also exist the extra degeneracy corresponding to the number of degenerate potential minima, for which the energy spacing closes exponentially as $g$ is increased. 
}
\end{figure*}

While significant efforts have been devoted to elucidating the strong and ultrastrong coupling regimes in the last decade \cite{ZH10,KK12,GT13,CDE13,GTA13,PB13,MR14,SBE14,PH15,CG16,TS16,FDP17,TS18,GN18,MJP19,SBE19,LS19,RRJ20,MS20,GGC21,FP19,KAF19,SAS21,RD11}, the physics of waveguide QED in the realms of even stronger light-matter interactions, namely, the deep \cite{CJ10} and extremely strong \cite{YA21}  coupling regimes, remains largely unexplored. 
There, the coupling strength becomes comparable to elementary excitation energies or exceeds them, and qualitatively different phenomena are expected to occur since vacuum fluctuations alone can lead to large populations of photons in every coupled mode. However, due to the aforementioned difficulties, a reliable theoretical approach for unveiling these intriguing phenomena is currently lacking. The primary goal of this paper is to reveal physics of strongly interacting light-matter systems in the previously unexplored regimes on the basis of a comprehensive theoretical framework that avoids problems discussed above.

On another front, the spin-boson model, a supposedly effective description of waveguide QED systems (e.g., Ref.~\cite{PB13}), has long been known to  exhibit the delocalization-localization transition at strong couplings \cite{BM70,CAO81}. Nevertheless, the breakdown of the usual two-level description \cite{YA21,DB182} and the relevance of the diamagnetic term \cite{DB182,RK75,NP10,VO11,AGM19,SA19,DLS14,GRJ15,YA21} have made it unclear until now how these known results for quantum dissipative systems should be interpreted in the context of waveguide QED. More specifically, the conditions under which a counterpart of the delocalization-localization transition exists should be carefully examined by using the full-fledged QED Hamiltonian. One intriguing possibility is that such a quantum phase transition can be extended to multi-emitter systems and provide a new route toward realizing a superradiant transition without external driving \cite{YA20,PP20,LS21}.

In view of recent experimental developments in realizing stronger light-matter interactions \cite{FDP17,YF172,MNS20}, the time is ripe to explore in and out of equilibrium physics of nonperturbative waveguide QED in a comprehensive manner. Specifically, we will ask the following questions:
\begin{itemize}
\item[(A)]{What are the defining physical features of waveguide QED in the previously unexplored nonperturbative regimes?}
\item[(B)]{How can one construct a proper effective model of waveguide QED at strong couplings, where existing theoretical descriptions are expected to fail?}
\item[(C)]{Starting from a fully microscopic theory, is it possible to identify a quantum phase transition akin to the delocalization-localization transition in waveguide QED setups, and if so, does there exist a new feature?}
\end{itemize}
The main aim of this paper is to reveal the new physics and develop understanding of strongly interacting light-matter systems by addressing these questions from a unified perspective.  Below we summarize the main results at a nontechnical level before presenting a detailed theoretical formulation in subsequent sections.

\subsection{Summary of the main results}
Our first main result is the appearance of a ladder of many-body  bound states (BS) and the many-body bound states in the continuum (BIC) in nonperturbative regimes (see Fig.~\ref{fig_schem}). We point out formation of increasingly many low-lying bound states whose energies decrease as $\propto\! g^{-1}$ in the limit of strong light-matter coupling $g$. The exact BICs emerge as a consequence of the $\mathbb{Z}_2$-symmetry that is linked to microscopic QED Hamiltonians. 
Previous studies have so far discussed the realizations of one-body BICs, which relied on artificial tuning of either emitter positions or resonator wavelengths/geometry \cite{TT13,CGB13,PF16,CG19,DF19,BK21}. In contrast, a new type of BICs found here does not rely on either of them, but emerge from strong light-matter interaction (without artificial fine tunings) and thus have a many-body origin.   
Even when the symmetry is not exact, the lifetime of these states diverges as $\propto\!g^{3/2}$ in the strong-coupling limit, and thus they still behave as so-called quasi BIC (see Fig.~\ref{fig_sp} in Sec.~\ref{sec:app}).

We note that the BICs have recently attracted significant attention in light of their potential applications for realizing quantum memory \cite{LAI09} and nondissipative emitter interactions \cite{GTA15,DJS15}. 
It is in general challenging to detect BICs in standard photon scattering experiments, since bound states are orthogonal to delocalized states in the continuum. Instead, we propose and numerically demonstrate an experimentally feasible quench protocol to excite the states in a model of cavity array \cite{HMJ06,ZL08,LF14}, leading to rich nonequilibrium dynamics in which the bound states and the dynamical Casimir effect are intertwined (see Fig.~\ref{fig_quench} in Sec.~\ref{sec:app}). 
These results establish one of the defining features in the nonperturbative regimes of waveguide QED and thus address question (A).

We remark that the present work should be contrasted to earlier studies of atom-field dressed bound states \cite{JS90} in several crucial aspects.  In Refs.~\cite{LP10,LP11,CG16,TS16,KS16,SMP16,BM19,MS20,LL21,KE21}, the existence of bound states was predicted in perturbative regimes on the basis of the rotating wave approximation. However, it turns out that these bound states in general become resonances with finite lifetimes once the counter rotating terms are included \cite{SBE14,RRJ20}. In contrast, our analysis does not rely on  those simplifying approximations and rigorously establishes the presence of bound states at arbitrary coupling strengths for general photonic dispersions. In particular, a ladder of bound states or BICs revealed by our analysis are appreciable only after multilevel structure of emitters is consistently included in theory.  We also remark that these bound states are genuine quantum many-body states in contrast to one-body wave phenomena, which have been the main focus of earlier studies \cite{HCW16}.

The second important result of our work is construction of proper effective models for waveguide QED that remain valid at arbitrary coupling strengths. This is made possible through the use of a unitary transformation that achieves asymptotic decoupling of emitter and photon degrees of freedom in the limit where light-matter interaction becomes the dominant energy scale (see Fig.~\ref{fig_schem}(a)).  We point out that conventional descriptions become inapplicable in the nonperturbative regimes because of uncontrolled level truncations in the Coulomb or Power-Zienau-Woolley (i.e., dipole) gauges.  In contrast, following the unitary transformation used in the present work, such truncations are  well-justified owing to vanishingly small light-matter entanglement at strong couplings,  ensuring the validity of effective models constructed in this new frame of reference. The obtained effective models take the same standard forms as the Jaynes-Cummings-type Hamiltonian for a single-emitter case (see Eq.~\eqref{rwaad} in Sec.~\ref{sec:app}) and the inhomogeneous transverse-field Ising Hamiltonian for a multi-emitter case (see Eq.~\eqref{ising} in Sec.~\ref{sec:cas}), but with suitably renormalized parameters.  These results address question (B).

Finally, building on these analyses, we answer question (C) in the affirmative way. Specifically, we show that the infrared divergence of the renormalized emitter mass occurs for a certain gapless photonic dispersion. This in turn implies the exact two-fold  degeneracy of the ground state and thus leads to the transition to the symmetry-broken (i.e., localized) phase in the thermodynamic limit, which is reminiscent of the delocalization-localization transition in the spin-boson models. Our results also indicate a qualitatively new feature, not present in the simplified spin-boson descriptions, such as  the reentrant transition into the delocalized phase in the extremely strong coupling regimes which originates from the mass acquisition in the transformed frame (see Figs.~\ref{fig_phase} and \ref{fig_frg} in Sec.~\ref{sec:gless}). We demonstrate these results by applying the functional renormalization group method to a concrete model of resistively shunted Josephson junctions. 

Overall, it is notable that the key features revealed by this paper do not rely on fine-tuning of parameters, but should appear generally in strongly coupled light-matter systems. To obtain these results, it is crucial to accurately perform analysis without resorting to uncontrolled approximations that cannot be justified in the nonperturbative regimes.
Below we thus start by developing a rigorous framework for describing  quantum emitters coupled to arbitrary multiple quantum electromagnetic modes, including the case of a continuum spectrum. This is done by extending the asymptotic light-matter decoupling unitary transformation that we introduced earlier in the context of single-mode cavity QED \cite{YA21} to the present waveguide QED setups. Most of the previous studies  approximated an emitter as a simplified two-level system, which, however, is not a valid approximation for many experimentally relevant systems, including superconducting qubits as mentioned before \cite{HMD17,ESS20,MWL20,RP20}. 
The validity of the two-level approximation becomes particularly questionable in the nonperturbative regimes due to significant renormalization of both the effective mass and potential as we demonstrate later (see e.g., Fig.~\ref{fig_rwa} in Sec.~\ref{subsec:two}). 
To provide an adequate model of the multilevel structure in realistic physical systems,  in the present work we model a quantum emitter as a charged particle moving in a potential with two degenerate minima (see Fig.~\ref{fig_schem}(a)).

While the emphasis of our discussion is on the waveguide setups, the present  formalism can be extended to other electromagnetic environments in arbitrary geometries. Examples include cavity QED systems in 2D materials  or polaritonic chemistry, in which the inclusion of multiple photonic modes becomes crucial depending on the cavity geometry and the coupling strength.  
Our work thus establishes a foundation for studying strongly coupled light-matter systems lying at the intersection of quantum optics, condensed matter physics, and quantum chemistry in genuinely nonperturbative regimes.

The remainder of the paper is organized as follows. In Sec.~\ref{sec:lm}, we present a general theoretical framework for a quantum emitter coupled to electromagnetic continuum on the basis of the asymptotically decoupling unitary transformation. In Sec.~\ref{sec:gen}, we unravel key physical features emerging in nonperturbative regimes of waveguide QED. In Sec.~\ref{sec:app}, we illustrate the general properties by providing an explicit numerical solution of a concrete model of coupled cavity arrays. In Sec.~\ref{sec:cas}, we present the extension of the theoretical formalism to multi-emitter systems. In Sec.~\ref{sec:gless}, we consider the ground-state properties of waveguide QED systems with a gapless photonic dispersion and discuss their relation to the delocalization-localization transition. In Sec.~\ref{sec:sum}, we give a summary of results and suggest several interesting directions for future investigations.

\section{Asymptotic Light-Matter Decoupling: general formalism\label{sec:lm}}
We first develop a general theory of a single quantum emitter coupled to arbitrary quantized electromagnetic environment. We use a disentangling unitary transformation that can asymptotically decouple light and matter degrees of freedom in the strong-coupling limit. This  {\it asymptotically decoupled} (AD) frame significantly simplifies the analysis of strongly interacting light-matter systems, which allows us to explore the entire coupling region, even beyond the ultrastrong coupling regimes.  We will later apply the framework to a concrete model of coupled cavity array in Sec.~\ref{sec:app}.  While a single-emitter setup is considered in this section, we will generalize the whole formalism to multi-emitter cases in Sec.~\ref{sec:cas}.

\subsection{QED Hamiltonian in the Coulomb gauge}
We consider a quantum emitter that is locally coupled to quantized electromagnetic modes in arbitrary geometries. The emitter is modeled as a quantum particle of mass $m$ and charge $q$ trapped by a potential $V$, while the electromagnetic environment is represented as a sum of harmonic oscillators with frequencies $\omega_k$. The corresponding QED Hamiltonian in the Coulomb gauge is given by 
\eqn{\label{HC}
\hat{H}_{{\rm C}}=\frac{\bigl(\hat{P}-q\hat{A}\bigr)^{2}}{2m}+V(\hat{Q})+\sum_{k}\hbar\omega_{k}\hat{a}_{k}^{\dagger}\hat{a}_{k},
}
where $\hat{Q}$ ($\hat{P}$) is the position (momentum) operator of the emitter and $\hat{a}_k$ ($\hat{a}_k^\dagger$) is the annihilation (creation) operator of photons in mode $k$, which satisfy the commutation relations
\eqn{
[\hat{Q},\hat{P}]=i\hbar,\;\;[\hat{a}_{k},\hat{a}_{k'}^{\dagger}]=\delta_{kk'}.
}
We denote the vector potential operator as 
\eqn{\label{vecpot}
\hat{A}=\sum_{k}f_{k}(\hat{a}_{k}+\hat{a}_{k}^{\dagger}),
}
where $f_k$ characterizes the electromagnetic amplitude of mode $k$. 

It is useful to diagonalize the quadratic photon part of $\hat{H}_{\rm C}$ as follows  (see Appendix~\ref{app_diag} for details):
\eqn{\label{HCb}
\hat{H}_{{\rm C}}\!=\!\frac{\hat{P}^{2}}{2m}+V(\hat{Q})\!-\!\hat{P}\sum_{n}\zeta_{n}\left(\hat{b}_{n}\!+\!\hat{b}_{n}^{\dagger}\right)\!+\!\sum_{n}\hbar\Omega_{n}\hat{b}_{n}^{\dagger}\hat{b}_{n},\nonumber\\
}
where we perform the canonical transformation to introduce a squeezed photon operator $\hat{b}_n$ labeled by $n\in\mathbb{Z}$ via
\eqn{\label{sqz}
\hat{a}_{k}=\sum_{n}\left(O\right)_{kn}\bigl[\cosh\left(r_{nk}\right)\hat{b}_{n}-\sinh\left(r_{nk}\right)\hat{b}_{n}^{\dagger}\bigr],
}
and $\zeta_n$ is given as
\eqn{
\zeta_{n}=\sqrt{\frac{\hbar}{m\Omega_{n}}}\sum_{k}g_{k}O_{kn}.
}
Here, $O_{kn}$ is an orthogonal matrix that satisfies 
\eqn{\label{orthogonal}
\sum_{kk'}\left(O^{{\rm T}}\right)_{nk}\left(\delta_{kk'}\omega_{k}^{2}+2g_{k}g_{k'}\right)O_{k'm}=\delta_{nm}\Omega_{n}^{2},
}
where $\Omega_n$ is an eigenfrequency of mode $n$,  $r_{nk}$ is a squeezing parameter defined by $e^{r_{nk}}\equiv\sqrt{\Omega_{n}/\omega_{k}}$,
and $g_k$ characterizes a coupling strength to mode $k$:
\eqn{\label{gkeq}
g_{k}\equiv qf_{k}\sqrt{\frac{\omega_{k}}{m\hbar}}.
}
We note that the magnitudes of $g_k$ depend on the size of the electromagnetic environment $L$ via $g_{k}\propto f_{k}\propto L^{-1/2}$. In a concrete model discussed later (cf. Eq.~\eqref{eqcca}), the environment consists of the coupled cavity arrays and the variable $L$ corresponds to the total number of cavities.

Before proceeding further, we make two remarks. 
First, while we follow the standard notation in atomic QED to write down the Hamiltonian~\eqref{HC}, the present formulation is equally applicable to circuit QED setups regardless of the physical nature of each variable. In superconducting circuits, artificial atoms are locally coupled to the continuum of microwave electromagnetic fields in a transmission line. There is a well-established analogy between circuit and atomic QED systems; the charge number operator of a transmon qubit and its conjugate phase operator precisely correspond to $\hat{P}$ and $\hat{Q}$ in Eq.~\eqref{HC}, respectively, and the charge bias induced by the electromagnetic fields of microwave resonator plays the role of the vector potential $\hat{A}$ (see also Sec.~\ref{sec:frg}). The same analogy holds true also for a flux qubit, where $\hat{Q}$ is coupled to photons through the dipole-type coupling $\hat{Q}\cdot\hat{E}$ with $\hat{E}$ being the electric field; one can use the Power-Zienau-Woolley (PZW) transformation \cite{PZ59,WRG71} to change this circuit Hamiltonian back to the standard form as in  Eq.~\eqref{HC} (see e.g., Ref.~\cite{DB182} or Eq.~\eqref{Hpzw} below). In practice, coefficients of the $\hat{A}^{2}$ term in circuit setups may have to be modified depending on resonator geometries. Our formalism below can readily be generalized to include such specifics.

{\renewcommand{\arraystretch}{1.5}
\begin{table*}
\caption{Summary of the scaling analysis for each of the renormalized parameters at different coupling strengths $g$ in Eq.~\eqref{gdef} normalized by the characteristic photon frequency $\omega$ in Eq.~\eqref{pome}.  The second, third, and fourth columns represent the results in the ultrastrong coupling (USC), deep strong coupling (DSC), and extremely strong coupling (ESC) regimes, respectively. The renormalized frequencies $\Omega_n$ are determined by the eigenvalue problem~\eqref{orthogonal}. The label $n\!=\!0$ indicates the dominant electromagnetic mode with the largest eigenfrequency. The  length scales $\xi_n$ in Eq.~\eqref{xin} are normalized by $x_\omega$ in Eq.~\eqref{xnorm} and characterize the effective light-matter coupling strengths for the dominant $n\!=\!0$ and the other modes $n\!\neq\!0$ in the transformed frame. The effective mass $m_{\rm eff}$ is defined by Eq.~\eqref{meff} with the renormalization factor~\eqref{mefftheta}. The two lowest rows correspond to the expectation values of the total photon numbers with respect to the low-energy eigenstates in the transformed frame (denoted by $U$) or in the Coulomb gauge (denoted by ${\rm C}$); see Eqs.~\eqref{photonAD} and \eqref{photonC}.}\label{table1}
\begin{tabular}{@{\hspace{8mm}} l @{\hspace{20mm}} l @{\hspace{30mm}} l @{\hspace{28mm}}l @{\hspace{20mm}}}
\hline
\hline
\multirow{2}{*}{Parameter} &\multicolumn{1}{l}{USC} &  \multicolumn{1}{l}{DSC} & \multicolumn{1}{l}{ESC}\\
&\multicolumn{1}{l}{$g/\omega\sim 0.1$}&\multicolumn{1}{l}{$g/\omega\sim 1$}&\multicolumn{1}{l}{$g/\omega>1$}\\[2pt] \hline
$\Omega_0/\omega$ & $\simeq 1$ & $\simeq\sqrt{1+2g^2/\omega^2}$ & $\propto g$  \\
$\Omega_{n\neq0}/\omega$ & $\simeq 1$ &  $\simeq 1$ & $\propto g^0$  \\
$\xi_{0}/x_\omega$ & $\simeq g^2/\omega^2$ & $\simeq\sqrt{\frac{g^{2}/\omega^{2}}{\left(1+2g^{2}/\omega^{2}\right)^{3/2}}}$ & $\propto g^{-1/2}$  \\
$\xi_{n\neq0}/x_\omega$ & $\simeq g/(\omega\sqrt{L})$ & $={\rm O}(\delta^2/(g\omega\sqrt{L}))$ & $\propto g^{-1}$  \\
$m_{\rm eff}/m$ & $\simeq 1$ & $\simeq 1+2g^2/\omega^2$ & $\propto g^2$  \\
$\langle\sum_{n}\hat{b}_n^\dagger\hat{b}_n\rangle_U$ & $={\rm O}(g^2/\omega^2)$ & $\simeq \frac{g^{2}/\omega^{2}}{\left(1+2g^{2}/\omega^{2}\right)^{5/2}}$ & $\propto g^{-3}$, $g^{-2}$  \\
$\langle\sum_{k}\hat{a}_k^\dagger\hat{a}_k\rangle_{\rm C}$ & $={\rm O}(g^2/\omega^2)$ & $={\rm o}(g^2/\omega^2)$ & $\propto g$  \\[3pt]
\hline
\hline
\end{tabular}
\end{table*}

Second, we invoke neither the two-level approximation of an emitter nor the rotating wave approximation (RWA), which are often used in the literature but will break down when the light-matter interaction becomes sufficiently strong. In particular, it will be crucial to take into account the multilevel structure of an emitter to unveil the key physics in nonperturbative regimes as we demonstrate later. We also note that the $\hat{A}^2$ term must be incorporated  to retain the gauge invariance of the theory, and its inclusion becomes particularly essential when one goes beyond the ultrastrong coupling regime. Meanwhile, we assume that the length scale of a quantum emitter is much smaller than the photon wavelength in such a way that the $\hat{Q}$ dependence of the vector potential $\hat{A}$ can be neglected. This long-wave assumption ultimately puts an upper limit on the light-matter coupling when the confinement length scale of the emerging localized mode analyzed below becomes comparable to the emitter size.

\subsection{Asymptotic decoupling transformation\label{subsec:asy}}
We now introduce a unitary transformation to asymptotically decouple light and matter degrees of freedom \cite{YA21}:
\eqn{\label{unitary}
\hat{U}=\exp\left(-\frac{i}{\hbar}\hat{P}\,\hat{\Xi}\right),
}
where $\hat{\Xi}$ is given as
\eqn{\label{xiope}
\hat{\Xi}&\equiv&\sum_{n}i\xi_{n}(\hat{b}^\dagger_n-\hat{b}_n),\\
\xi_{n}&\equiv&\frac{\zeta_{n}}{\Omega_{n}}.\label{xin}
}
This transformation acts on individual operators via
\eqn{
\hat{U}^{\dagger}\hat{Q}\hat{U}&=&\hat{Q}+\hat{\Xi},\\
\hat{U}^{\dagger}\hat{b}_{n}\hat{U}&=&\hat{b}_{n}+\frac{\xi_{n}\hat{P}}{\hbar},\label{bshift}
}
where the emitter position is shifted by the gauge-field-dependent displacement $\hat{\Xi}$ while each photon mode is subject to the momentum-dependent shift $\xi_n\hat{P}/\hbar$ \footnote{We remark that the field $\hat{\Xi}$ here should not be confused with the quantity referred to as the displacement field, $\hat{D}=\epsilon_0\hat{E}\!+\!\hat{P}$, with polarization $\hat{P}$ in the context of macroscopic electrodynamics.}.  We note that the displacement variables $\xi_n$ in Eq.~\eqref{xin} are chosen in such a way that the $\hat{P}\!\cdot\!(\hat{b}\!+\!\hat{b}^\dagger)$ term in Eq.~\eqref{HCb} will be precisely cancelled by the contributions arising from the displacement of the $\hat{b}^\dagger\hat{b}$ term via Eq.~\eqref{bshift}.

The resulting Hamiltonian in the asymptotically decoupled frame is
\eqn{\label{HU1}
\hat{H}_{U}&=&\hat{U}^{\dagger}\hat{H}_{{\rm C}}\hat{U}\nonumber\\
&=&\frac{\hat{P}^{2}}{2m_{{\rm eff}}}+V(\hat{Q}+\hat{\Xi})+\sum_{n}\hbar\Omega_{n}\hat{b}_{n}^{\dagger}\hat{b}_{n}.
}
Here the effective mass is defined as 
\eqn{
m_{{\rm eff}}&\equiv& m(1+2\Theta),\label{meff}\\
\Theta&\equiv&\sum_{k}\left(\frac{g_{k}}{\omega_{k}}\right)^{2},\label{mefftheta}
}
where the mass enhancement is characterized by the dimensionless quantity $\Theta$ whose expression~\eqref{mefftheta} follows from  Eq.~\eqref{orthogonal}. This renormalization comes from the $\hat{P}^2$ terms arising from the residual contributions generated by displacing the $\hat{P}\!\cdot\!(\hat{b}\!+\!\hat{b}^\dagger)$ and  $\hat{b}^\dagger\hat{b}$ terms. 
After the transformation, the light-matter interaction is incorporated in the external potential $V$ in the form of the gauge-field-dependent shift of the emitter, and its effective coupling strength is characterized by $\xi_n$ instead of the bare coupling $g_k$. 

Hereafter we first focus on the case of a gapped dispersion with frequencies $\omega_k\!>\!0$  $ \forall k$, for which $\Theta$ remains finite. This includes experimentally relevant systems such as coupled cavity arrays and open microwave transmission lines.  The case of a gapless dispersion should be analyzed separately, since one can find an infrared divergence of $\Theta$ in that case; we will revisit this issue in Sec.~\ref{sec:gless}. 

\subsection{Scaling analysis of the effective parameters}
To demonstrate the asymptotic light-matter decoupling, we perform the scaling analysis of the renormalized parameters with respect to the interaction strength. To this end, we introduce the characteristic photonic frequency $\omega$ and the coupling strength $g$ as follows: 
\eqn{\label{pome}
\omega^{2}&\equiv&\sum_{k}\omega_{k}^{2}/L,\\
g^{2}&\equiv&\sum_{k}g_{k}^{2},\label{gdef}
}
where we note $g\!=\!{\rm O}(L^0)$ since $g_k\!\propto\! L^{-1/2}$.
We begin by considering the regime $g/\omega\!>\! 1$ in which the coupling strength is dominant over other energy scales; we shall refer to it as the extremely strong coupling (ESC) regime \cite{YA21}. There, the eigenvalue problem~\eqref{orthogonal} has a single dominant mode (which we label $n\!=\!0$) with the largest eigenfrequency $\Omega_0\!\propto\! g$ and $\sum_k g_kO_{k0}\!\simeq\! g$, while the other frequencies remain $\Omega_{n\neq 0}\!=\!{\rm O}(\omega)$ with $\sum_k g_kO_{kn\neq0}\!\propto\!{\rm O}(g^{-1})$. This leads to the scalings $\xi_0\propto g^{-1/2}$ and $\xi_{n\neq0}\propto g^{-1}$, i.e., the  light-matter interaction in $\hat{H}_U$ asymptotically vanishes in the strong-coupling limit.

In the deep strong coupling (DSC) regime $g/\omega\!\sim\! 1$ \cite{CJ10}, one can continue the scaling analysis in the similar manner and obtain slightly refined  expressions as summarized in Table~\ref{table1}. There, we denote the variance of a photonic dispersion (or the effective bandwidth) as 
\eqn{\delta^{2}=\sum_{k}(\omega_{k}-\omega)^{2}/L,\label{dvariance}} 
and normalize the length scale by the characteristic one
\eqn{x_{\omega}=\sqrt{\frac{\hbar}{m\omega}}.\label{xnorm}} 
In Table~\ref{table1}, we also summarize the scaling relations in the ultrastrong coupling (USC) regime $g/\omega\sim0.1$, which can readily be obtained by the perturbative analysis.
All these scalings will later be demonstrated in a case study of coupled cavity array  (see Fig.~\ref{fig_xi} below).

Besides the asymptotic decoupling in the strong-coupling limit, 
one notable result of this scaling analysis is that the displacement parameters $\xi_n$ and consequently the effective light-matter couplings in the AD frame remain small over the entire region of $g$. As detailed below, this fact allows us to significantly simplify the analysis in a broad range of coupling strengths, including the realms beyond the USC regime which are otherwise challenging to investigate in any previous theoretical approaches.

\subsection{Vacuum-dressed potential and decoupled excitations}
To analyze low-energy eigenstates of $\hat{H}_U$, it is useful to rewrite it in the following manner:
\eqn{
\hat{H}_{U}=\hat{H}_{{\rm matter}}+\hat{H}_{{\rm int}}+\hat{H}_{{\rm light}},
}
where we define the matter Hamiltonian by
\eqn{\label{Hmat}
\hat{H}_{{\rm matter}}=\frac{\hat{P}^{2}}{2m_{{\rm eff}}}+V_{{\rm eff}}(\hat{Q})
}
with $m_{\rm eff}$ being the renormalized mass~\eqref{meff} and $V_{\rm eff}$ being the dressed potential given as
\eqn{\label{Veff}
V_{{\rm eff}}(Q)&\equiv& V(Q)+\sum_{l=1}\frac{\xi^{2l}}{(2l)!!}V^{(2l)}(Q),\\
\xi^{2}&\equiv&\sum_{n}\xi_{n}^{2},\label{xiveff}
}
where $V^{(l)}$ is the $l$-th derivative of $V$. 
The interaction Hamiltonian is given by
\eqn{\label{intno}
\hat{H}_{{\rm int}}=\sum_{l=1}\frac{:\hat{\Xi}^{l}:}{l!}V^{(l)}(\hat{Q}),
}
where $:\hat{O}:\equiv\hat{O}-\langle0|\hat{O}|0\rangle$ represents the normal ordering of photonic operators with $|0\rangle$ being the vacuum state in the AD frame:
\eqn{
\hat{b}_n|0\rangle=0\;\;\;\forall{n}.
}
We emphasize that this vacuum state is distinct from the original vacuum of $\hat{a}$ operators in the Coulomb gauge due to the squeezing (cf. Eq.~\eqref{sqz}). 
Finally, we denote the photon Hamiltonian as 
\eqn{\label{photonham}
\hat{H}_{{\rm light}}=\sum_{n}\hbar\Omega_{n}\hat{b}_{n}^{\dagger}\hat{b}_{n}.
}
Equations~\eqref{Veff} and \eqref{intno} can be obtained by expanding the interaction term in Eq.~\eqref{HU1} with respect to $\hat{\Xi}$ and using the relations $\langle0|\hat{\Xi}^{2l}|0\rangle\!=\!(2l-1)!!\xi^{2l}$ and $\langle0|\hat{\Xi}^{2l-1}|0\rangle\!=\!0$.

Arguably, the most celebrated signature of quantum fluctuations of the electromagnetic field for an isolated ground-state atom is the Lamb shift. Incorporation of light-matter coupling exclusively through a modification of the external potential $V(\hat{Q}\!+\!\hat{\Xi})$ renders the origin of the Lamb shift explicit; fluctuations in $\hat{\Xi}$ add to the intrinsic fluctuations of $\hat{Q}$ to enhance the variance of the effective particle position. This feature manifests itself as the nonvanishing dressing in the effective potential~\eqref{Veff} without any externally excited photons; it originates from the zero-point fluctuations of the electromagnetic fields. Here, the dominant contribution to the dressing strength $\xi$ comes from the mode $n=0$ and thus $\xi$ basically obeys the same scaling relation satisfied by $\xi_0$ in Table~\ref{table1}. If necessary, the summation over $l$ can in practice be truncated at a certain order that scales inversely with the coupling strength $g$ owing to the asymptotic vanishment of $\xi$.
In particular, in the limit of weak light-matter coupling, the energy shift of the ground state is captured by the $l\!=\!1$ term of Eq.~\eqref{Veff} and the effective-mass enhancement in Eq.~\eqref{meff}.

Because of the  decoupling $\xi_0\!\propto\! g^{-1/2}$ and 
enhancement of the effective photon frequency $\Omega_0\!\propto\! g$, low-energy eigenstates of $\hat{H}_U$ in the strong-coupling limit can be written as a product of the emitter eigenstates and the photon vacuum as follows:
\eqn{\label{deceig}
|\Psi_\alpha\rangle_U=|\psi_\alpha\rangle|0\rangle,
}
where $|\psi_\alpha\rangle$ with $\alpha=1,2,\ldots$ are single-particle eigenstates of $\hat{H}_{\rm matter}$ in Eq.~\eqref{Hmat};  we then represent it as 
\eqn{
\hat{H}_{\rm matter}=\sum_{\alpha}E_{\alpha}(g)|\psi_{\alpha}\rangle\langle\psi_{\alpha}|
}
with $E_{1}\!\leq\! E_{2}\!\leq\!\cdots$ being the corresponding eigenenergies. 
These single-particle energies $E_\alpha$ provide asymptotically exact excitation energies of the total Hamiltonian $\hat{H}_U$, which in the original Coulomb gauge corresponds to an intrinsically many-body problem with highly entangled light-matter degrees of freedom (cf. Eq.~\eqref{HC}). Said differently,   when transforming back to the original Coulomb gauge, the above decoupled emitter states are in general entangled, light-matter correlated states.  We note that the mass enhancement $m_{\rm eff}\propto g^2$ leads to the tight localization of $|\psi_\alpha\rangle$ around the bottom of $V_{\rm eff}$; accordingly,  the excitation energies $\delta E_\alpha\equiv E_\alpha -E_1$ decrease as $\delta E_\alpha\propto g^{-1}$ in the nonperturbative regimes as long as $V_{\rm eff}$ has well-defined minima.

\subsection{Few-photon ansatz at arbitrary coupling strengths}
The scaling analysis indicates that the total number of photons in the AD frame remains small over the entire coupling region (cf. Table~\ref{table1}). In particular, in the ESC regime, the standard perturbation theory predicts that the photon number in the ground state is on the order of $(\xi_0/\Omega_0)^2$  as long as the bandwidth $\delta$ is narrow, resulting in the scaling
\eqn{\label{photonAD}
\bigl\langle\sum_{n}\hat{b}^\dagger_n\hat{b}_n\bigr\rangle_U\propto g^{-3},
}
where $\langle\cdots\rangle_U$ represents an expectation value with respect to a low-energy eigenstate in the AD frame.
As the bandwidth becomes broad, the contributions from $n\neq0$ modes eventually dominate the $n\!=\!0$ contribution above, and the perturbation theory leads to $\sum_{n\neq0}(\xi_n/\Omega_n)^2\!\propto\! g^{-2}$, where we used $\xi_{n\neq0}\!\propto\!\delta^2/(g\omega\sqrt{L})$ and $\Omega_{n\neq 0}\sim \omega$ (see Table~\ref{table1}). This crossover occurs when the bandwidth reaches a value around $g\delta^4/\omega^5={\rm O}(1)$, at which the contributions from $n\!\neq\!0$ and $n\!=\!0$ modes become comparable. In any case, the total photon number in the transformed frame asymptotically vanishes in the strong-coupling limit.

This fact motivates us to introduce the {\it few-photon ansatz}  by projecting the whole Hilbert space onto the following subspace:
\eqn{\label{fewphoton}
{\cal H}_{N_c}\equiv{\rm span}\left\{|\psi_{\alpha}\rangle\otimes|\psi_{{\rm photon},i}\rangle\right\},
}
where we recall that $|\psi_\alpha\rangle$ are single-particle eigenstates of $\hat{H}_{\rm matter}$ in Eq.~\eqref{Hmat} while we denote $|\psi_{{\rm photon},i}\rangle$ as an arbitrary bosonic many-body state that satisfies
\eqn{
\langle\psi_{{\rm photon},i}|\sum_{n}\hat{b}_{n}^{\dagger}\hat{b}_{n}|\psi_{{\rm photon},i}\rangle\leq N_{c}
}
with a photon-number cutoff $N_{c}$; note that this cutoff is imposed on the {\it total} photon number summed over all the modes, but not on each individual electromagnetic mode. 
With this definition, the decoupled excitations~\eqref{deceig} discussed above correspond to the simplest subspace ${\cal H}_{0}$ with no photons.

We here emphasize that the complexity is no longer exponential, but it is polynomial with respect to the system size $L$; the Hilbert-space dimension of the few-photon manifold ${\cal H}_{N_c}$ scales as $\propto L^{N_c}$. This allows us to study the (exact) waveguide QED Hamiltonian~\eqref{HC} at arbitrary coupling strengths in a very efficient and accurate manner. Indeed, our exact diagonalization analysis shows that the results converge within (at most) $\sim\!1$\% deviation already at a small total photon-number cutoff $N_c\!=\!2$-$4$ (see Appendix~\ref{app:num} for further details).

This point should be contrasted to previous approaches; eigenstates in the Coulomb gauge can possess large photon occupation numbers (see also Sec.~\ref{subsec:bre} below), and one has to include more excitations for each electromagnetic mode at a greater coupling $g$. Hence, the corresponding Hilbert-space dimension exponentially increases with $L$, which severely limits their applicabilities in the strong-coupling regions.  Some variational states, such as the displaced-oscillator  states \cite{SR84,BS14,GN18,RRJ20}, can provide useful approximative methods up to a rather modest coupling regime. However, they should also ultimately become inaccurate especially beyond the USC regime because of the breakdown of the polaron picture. More importantly, the usual two-level description of an emitter, on which most of the previous studies rely, becomes invalid once one enters into the DSC and ESC regimes as detailed below. We will show that it is actually such multilevel structure that leads to a defining feature of the waveguide QED in genuinely nonperturbative regimes.  
Our approach gets around these difficulties by employing the (asymptotically exact) disentangling unitary transformation, after which the whole low-energy eigenstates are well restricted into the few-photon manifold~\eqref{fewphoton} at any coupling strengths.

\section{Generic features of nonperturbative  waveguide QED \label{sec:gen}}
We now present key physical features of waveguide QED that emerge when one enters into the nonperturbative regimes on the basis of the theoretical formalism developed in Sec.~\ref{sec:lm}. To understand the qualitative physics, it is sufficient, as a first step, to consider the decoupled excitations~\eqref{deceig} that belong to the simplest, zero-photon subspace ${\cal H}_{0}$. The results discussed here establish universal nonperturvative features which hold true independent of specific choices or fine-tuning of  microscopic parameters. We will make these predictions quantitatively accurate in the next section by extending the analysis to the few-photon ansatz in  the subspace  ${\cal H}_{N_c>0}$.

\subsection{Ladder of many-body bound states}
One of the most surprising results is the appearance of a ladder of many-body bound states. To see this, we recall that  the excitation  energies of the decoupled states~\eqref{deceig} decrease as $\delta E_\alpha\!\propto\!1/g$ due to the mass enhancement, and eventually lie outside of the photon continuum,
\eqn{\label{excbs}
\delta E_\alpha\notin[\hbar\omega_{\rm L},\hbar\omega_{\rm U}],
}
where $\omega_{{\rm L}({\rm U})}$ represents the lower (upper) frequency limit of the photon dispersion. These states are energetically separated from scattering states and thus form bound states (BS), i.e., the excitation energies are localized to the emitter degree of freedom and cannot decay to the continuum at all. This emergence of multiple low-lying BS is inaccessible by the commonly used two-level treatments that can be valid only up to the USC regime. For this reason, the appearance of BS ladder can be considered as one of the defining features of the DSC and ESC regimes of waveguide QED.

 Interestingly, these bound states appear with equal energy spacing that narrows as $\delta E\propto 1/g$. This results from the increase of the emitter mass $m_{\rm eff}\!\propto\!g^2$ and the ensuing tight localization of the wavefunction, which can be best understood in the AD frame. The low-energy spectrum then reduces to the harmonic one as long as the potential is well-behaved and can be expanded quadratically around the minima.
Importantly, in the original frame, these states behave as the many-body BS, which are strongly entangled states including high-momentum emitter states and exponentially localized (virtual) photons. 
It is worthwhile to note that photon localization in these bound states becomes increasingly tight at greater $g$ and can be much smaller than the (bare) emitter-transition wavelength; this feature should be contrasted  to usual atom-field dressed bound states \cite{JS90}.

\subsection{Many-body bound states in the continuum\label{subsec:bic}}
Even when the decoupled excitations~\eqref{deceig} lie within the photon continuum, they  can behave as either symmetry-protected bound states in the continuum (BIC) or  quasi BIC with lifetime that diverges in the strong-coupling limit. 
 To understand the origin of such symmetry-protected BIC, suppose that the potential respects the inversion symmetry, $V(Q)\!=\!V(-Q)$. The total light-matter Hamiltonian then satisfies the following $\mathbb{Z}_2$ symmetry:
 \eqn{
\hat{{\cal P}}^{-1}\hat{H}_{U}\hat{{\cal P}}&=&\hat{H}_{U}, \label{z2sym}\\\hat{{\cal P}}^{2}&=&1,
 }
where $\hat{\cal P}$ acts on the emitter operators as $\hat{{\cal P}}^{-1}\hat{P}\hat{{\cal P}}=-\hat{P}$, $\hat{{\cal P}}^{-1}\hat{Q}\hat{{\cal P}}=-\hat{Q},$ and also transforms the photon field via  $\hat{{\cal P}}^{-1}\hat{b}_{n}\hat{{\cal P}}=-\hat{b}_{n}$. We emphasize that this $\mathbb{Z}_2$ symmetry is intrinsically  linked to microscopic light-matter Hamiltonians without making artificial fine tuning. For instance, in a circuit setup, the potential term $V(Q)$ typically consists of the sum of the Josephson energy $-E_J\cos(Q)$ and the inductive term $E_LQ^2/2$, both of which clearly satisfy the above symmetry. At a more fundamental level, since the first-principle QED Hamiltonian in the Coulomb gauge also naturally satisfies this symmetry \cite{CCT89}, the present QED Hamiltonian~\eqref{HC} should also respect that in general.

 It is then clear that, if a decoupled excitation lying in the photon continuum has a different parity from that of scattering states,  it leads to the exact BIC protected by the above $\mathbb{Z}_2$ symmetry. For instance, the lowest photon continuum has the odd parity, while a half of  the decoupled excitations~\eqref{deceig} have the even parity and thus can behave as the BIC when the excitation energies lie within the continuum.

Interestingly, even if a decoupled excitation has the same parity as scattering states, it can still behave as a long-lifetime resonance, which is often called quasi BIC. Indeed, the scaling analysis  of its decay rate given by the Fermi's golden rule results in
\eqn{\label{qbicdec}
\Gamma_{\rm qBIC}\propto g^{-3/2},
}
which vanishes in the strong-coupling limit. 
The same argument also applies to the case when the $\mathbb{Z}_2$ symmetry is not exact due to, e.g., the broken inversion symmetry, $V(Q)\!\neq\!V(-Q)$; the symmetry-protected BIC discussed above then become resonances in this case, however, their lifetimes still diverge in the strong-coupling limit.  Physically, these (quasi) BICs originate from the strong light-matter interaction containing the diamagnetic effect, which tends to prevent scattering photons from interacting with  the many-body BS consisting of virtual photons localized around the emitter. 
We emphasize that the physics of (quasi) BIC discussed here qualitatively remains the same also in the case of a gapless dispersion  unless the mass renormalization factor $\Theta$ diverges (see Sec.~\ref{sec:cas}).

It is worthwhile to note again that fine-tuning of the coupling strength is not necessary to observe the (quasi) BIC here. Specifically, there always exists a nonzero-measure parameter regime of $g$ such that a certain excitation lies in the photon continuum,
\eqn{\label{bicex}
\delta E_\alpha(g)\in[\hbar\omega_{\rm L},\hbar\omega_{\rm U}].
}
The reason for this is as follows. At zero coupling, one can find an emitter state lying above the continuum, i.e., $\delta E_\alpha(g\!=\!0)\!>\!\hbar\omega_{\rm U}$.  In the ESC regime, this excitation energy asymptotically decreases as $\delta E_\alpha(g)\!\propto\!g^{-1}$ and ultimately converges to zero. Thus, between these two limits, there must exist an intermediate coupling regime such that the relation~\eqref{bicex} is satisfied. The metastability of these states stems from the fact that radiation field modes that are resonant with them have small amplitudes at the emitter position. 

\subsection{Vacuum-induced suppression in potential barrier\label{subsec:vac}}
Yet another common feature in the nonperturbative regimes is the vacuum-induced suppression of potential barrier in $V_{\rm eff}$. 
This can readily be understood from Eq.~\eqref{Veff}, where the vacuum fluctuations decrease (increase) the energies at local maxima (minima), thus lowering the potential barrier felt by the particle when tunneling to a different local minimum (see Fig.~\ref{fig_veff} below for an illustrative example of the double-well potential). The amount of this suppression nonmonotonically depends on the coupling strength, since it is solely determined by the displacement parameter $\xi$ that exhibits the nonmonotonic $g$ dependence (see Eq.~\eqref{xiveff} and Table~\ref{table1}). 

When one considers quantum tunneling, the mass renormalization eventually dominates the barrier suppression and thus the tunneling rate is ultimately exponentially suppressed in the strong coupling limit. In contrast, if thermal activation over the barrier, i.e., thermal escape, is of interest, the escape rate is basically characterized by the ratio of the potential barrier to the temperature, but independent of the mass. Thus, it is a universal feature that a thermal escape should be {\it enhanced} by strong light-matter couplings owing to the vacuum-induced suppression of the barrier. This may provide a physical explanation of recent experimental observations in polaritonic chemistry \cite{Hiura2018,Hiura2019,TA19}, where the thermally activated chemical reaction  was found to be enhanced due to cavity confinement.

\subsection{Breakdown of level truncations in the Coulomb and PZW gauges\label{subsec:bre}}
We finally point out that an analysis relying on the Coulomb or PZW gauges must in general become invalid at a sufficiently strong light-matter coupling.  
This difficulty arises from the breakdown of level truncations in photon and emitter degrees of freedom in the nonperturbative regimes. Specifically, we first note that the mean photon number in the Coulomb gauge grows as (cf. Table~\ref{table1})
\eqn{\label{photonC}
\bigl\langle\sum_k\hat{a}^\dagger_k\hat{a}_k\bigr\rangle_{{\rm C}}\propto g.
}
The same scaling also applies to the photon-number fluctuations. 
 The number of photon basis required to analyze the Coulomb-gauge Hamiltonian~\eqref{HC} thus exponentially diverges as $g$ is increased. 
This eventually invalidates truncation of photon levels, which is actually inevitable in almost any analysis of bosonic many-body systems.

Similarly, the truncation of matter levels also becomes ill-justified at a sufficiently large $g$ in the conventional gauges; in particular, this fact indicates  the breakdown of the usual two-level descriptions of quantum emitters in the nonperturbative regimes. To see this, we  use the unitary transformation~\eqref{unitary} to express the decoupled states~\eqref{deceig} in the Coulomb gauge
\eqn{
|\Psi_{\alpha}\rangle_{\rm C}&=&\hat{U}|\psi_{\alpha}\rangle|0\rangle\nonumber\\
&=&\int dP\,\psi_{\alpha}(P)|P\rangle e^{-iP\hat{\Xi}}|0\rangle,
}
where we expand an emitter state $|\psi_\alpha\rangle$ in terms of the momentum eigenstates $|P\rangle$. In the strong-coupling limit, the variance of the momentum distribution $|\psi_\alpha(P)|^2$ diverges with $\sigma_P\!\propto\! g$ due to the mass renormalization $m_{\rm eff}\!\propto\! g^2$. Thus, an energy eigenstate in the Coulomb gauge is a strongly entangled emitter-photon state consisting of a superposition of higher momentum states at larger $g$. This fact eventually invalidates the common analyses that rely on either two-level approximation or a fixed momentum cutoff for an emitter.

Note that these difficulties are carried over in the PZW gauge (also known as the dipole gauge). To see this, we recall that the corresponding Hamiltonian in the long-wavelength limit  is given by $\hat{H}_{\rm PZW}\!=\!\hat{U}^\dagger_{\rm PZW}\hat{H}_{\rm C}\hat{U}_{\rm PZW}$ with the PZW transformation $\hat{U}_{\rm PZW}\!=\!\exp(iq\hat{Q}\hat{A}/\hbar)$:
\eqn{\label{Hpzw}
\hat{H}_{{\rm PZW}}&=&\frac{\hat{P}^{2}}{2m}+V(\hat{Q})+mg^{2}\hat{Q}^{2}+q\hat{Q}\hat{E}+\sum_{k}\hbar\omega_{k}\hat{a}_{k}^{\dagger}\hat{a}_{k},\nonumber\\
\hat{E}&\equiv&\sum_{k}if_{k}\omega_{k}(\hat{a}_{k}^{\dagger}-\hat{a}_{k}).
}
As is the case with the Coulomb gauge, the mean/fluctuation of the photon number in this gauge rapidly grows at strong couplings, while the mass remains at the bare value which leads to eventual breakdown of matter-level truncation. 
In contrast, the AD frame makes both photon- and emitter-level truncations well-justified and allows us to reveal the key features in the nonperturbative regimes as outlined above. We remark that, when transforming back to the Coulomb gauge, $\hat{E}$ in the above PZW gauge corresponds to the dielectric displacement field consisting of the electric field and the emitter shift.

\section{Application to coupled cavity array \label{sec:app}}
We here demonstrate all the generic features discussed in Sec.~\ref{sec:gen} by analyzing a concrete model of coupled cavity arrays. Extending the above analysis to the few-photon ansatz~\eqref{fewphoton}, we provide experimentally testable predictions of bound states, excitation energies, and quench dynamics, which are quantitatively accurate over the entire coupling region.  

\begin{figure}[b]
\includegraphics[width=86mm]{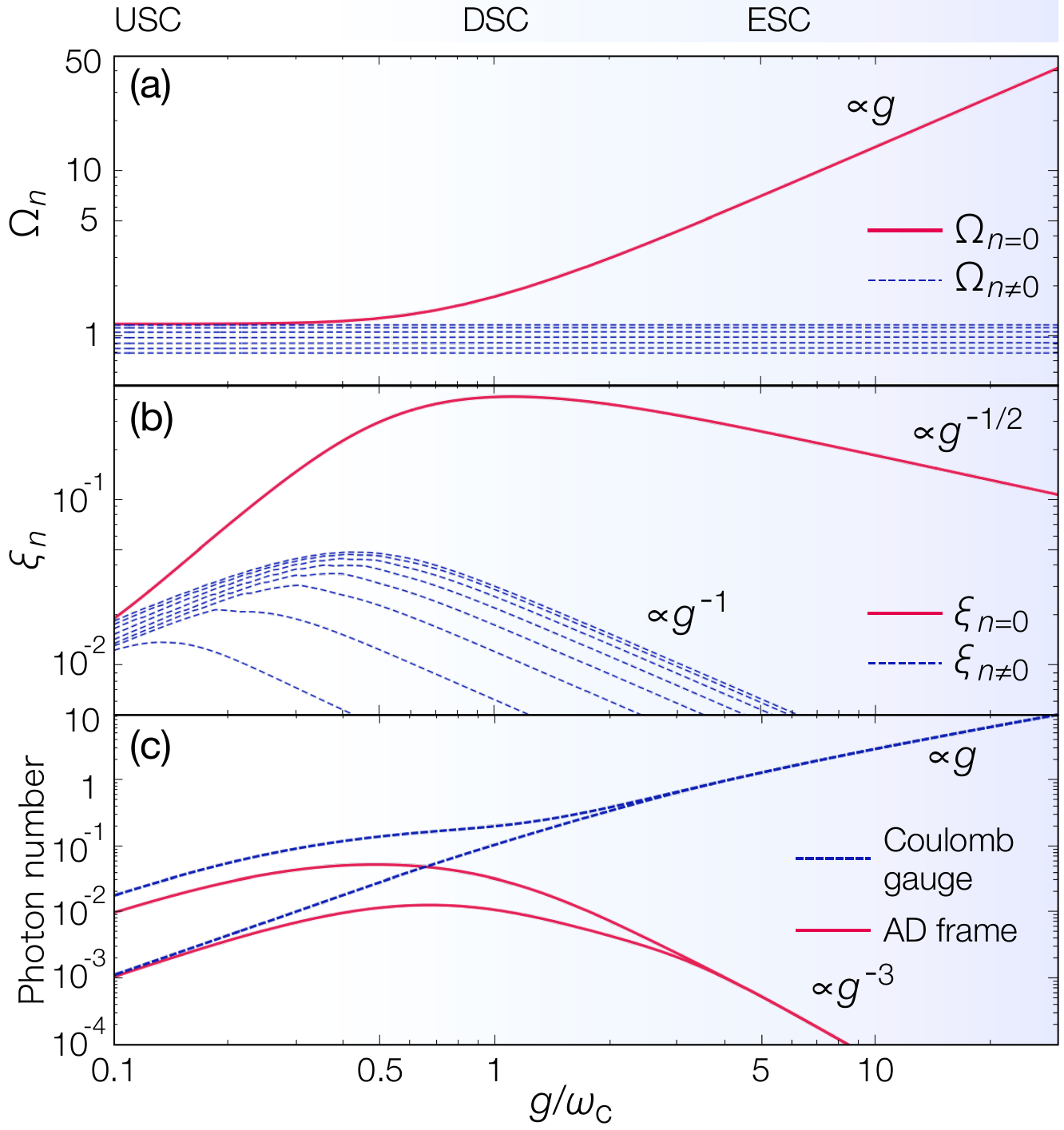} 
\caption{\label{fig_xi}
(a) Renormalized photon frequencies $\Omega_n$ in Eq.~\eqref{orthogonal}, (b) displacement parameters $\xi_n$ in Eq.~\eqref{xin}, and (c) expectation values of the total photon number plotted against the light-matter coupling strength $g$ in Eq.~\eqref{gcavity}. In (a,b), the red solid (blue dashed) curve shows the values corresponding to the dominant mode $n\!=\!0$ (the other modes $n\neq0$), where the dominant mode is characterized by the highest frequency $\Omega_{0}\!\simeq\!\sqrt{\omega_c^2+2g^2}$. Note that only a part of $n\!\neq\!0$ modes is plotted for the sake of visibility.  In (c), the red solid curves show the total number of dressed photons for the two-lowest eigenstates in the asymptotically decoupled (AD) frame (cf. Eq.~\eqref{photonAD}), while the blue dashed ones  show the corresponding number of bare photons in the Coulomb gauge (cf. Eq.~\eqref{photonC}).  Parameters are $J=0.2$ in (a,b),  and $J=0.2$, $v=0.5$, and $d=1.2$  in (c). 
}
\end{figure}

We consider a waveguide realized by a one-dimensional array of coupled cavities with nearest-neighbor coupling
\eqn{\label{eqcca}
\hat{H}_{{\rm light}}\!=\!-\frac{J}{2}\sum_{x}\left(\hat{a}_{x+1}^{\dagger}\hat{a}_{x}\!+\!{\rm H.c.}\right)\!+\!\hbar\omega_{c}\!\sum_{x}\hat{a}_{x}^{\dagger}\hat{a}_{x},
}
where $J$ is a hopping parameter, $\omega_c$ is a resonator frequency, and
$\hat{a}_{x}\equiv\frac{1}{\sqrt{L}}\sum_{k}\hat{a}_{k}e^{-ikx}$ is a photonic annihilation operator of the resonator mode at site $x$. The corresponding dispersion is
\eqn{
\hbar\omega_{k}=\hbar\omega_{c}-J\cos k
}
with wavevector $k\in[-\pi,\pi)$. This specific choice of the waveguide is not essential to the qualitative physics we discuss below, but is amenable to numerical calculations and experimental implementations. The emitter is coupled to the waveguide at $x\!=\!0$ and the vector potential in the Coulomb-gauge Hamiltonian~\eqref{HC} is given by
\eqn{
\hat{A}={\cal A}(\hat{a}_{x=0}+\hat{a}_{x=0}^{\dagger}),
}
which corresponds to electromagnetic amplitudes $f_k$ (see Eq.~\eqref{vecpot})
\eqn{\label{fcoupling}
f_{k}=\frac{{\cal A}}{\sqrt{L}}.
}
As the amplitudes are independent of $k$, it is useful to define the characteristic light-matter coupling strength by
\eqn{\label{gcavity}
g=q{\cal A}\sqrt{\frac{\omega_c}{m\hbar}}.
}
Note that this expression is consistent with the definition~\eqref{gdef}.

We model a quantum emitter as a charged particle trapped in the standard double-well potential,
\eqn{\label{doublewell}
V(Q)=v\left(1-\frac{Q^{2}}{d^{2}}\right)^{2},
}
where $v$ is a potential depth and $d$ characterizes a position of the potential minima. While we here focus on this minimal model for a quantum emitter, our theoretical formalism is equally applicable to a general potential landscape that may be more complex depending on each specific system or problem, such as transmon/flux qubits or chemical reactions.

Figure~\ref{fig_xi}(a,b) shows the renormalized parameters in the AD frame at different coupling strengths; all the numerical values are shown in the unit $\omega_c\!=\!\hbar\!=\!m\!=\!1$ throughout this paper.
These results are fully consistent with the scaling analysis summarized in Table~\ref{table1}. Specifically, beyond the USC regime, a single mode labeled by $n\!=\!0$ turns out to have a large renormalized eigenfrequency and dominantly couples to the emitter, while the other modes with $n\!\neq\!0$ basically remain at the bare frequencies and are almost decoupled from the emitter. As shown in  Fig.~\ref{fig_xi}(c), the total photon number in the AD frame vanishes as $\propto\!g^{-3}$ as consistent with the scalings $\Omega_0\!\propto\! g$ and $\xi_0\!\propto\! g^{-1/2}$, while in the Coulomb gauge the photon number increases as $\propto \!g$.

\subsection{Bound states, symmetry-protected BIC, quasi BIC}
\begin{figure*}[t]
\includegraphics[width=160mm]{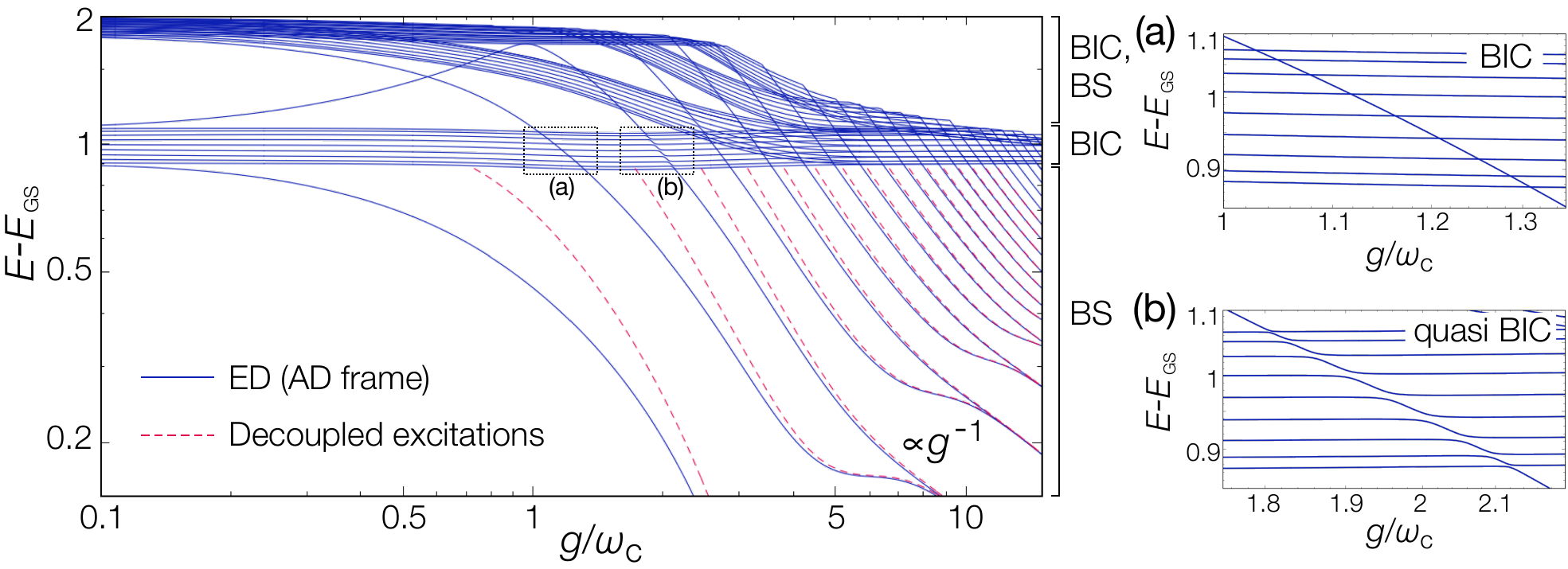} 
\caption{\label{fig_sp}
(Left) Low-energy excitation spectrum obtained by the exact diagonalization (ED) of the AD-frame Hamiltonian~\eqref{HU1} with the double-well potential~\eqref{doublewell} at different coupling strengths $g$. The red dashed curves show the energies corresponding to the decoupled excitations~\eqref{deceig}, which become asymptotically exact in the strong-coupling limit. Panels (a) and (b) are closeups of the main panel, where the exact BIC shows no anticrossings in (a), while the quasi BIC exhibits tiny anticrossings in (b).
Parameters are $J=0.1$, $v=0.5$, $d=0.87$, and $L=19$. The emitter parameters correspond to the on-resonant condition $\Delta/\omega_c\simeq1$ with $\Delta$ being the bare emitter frequency (see Eqs.~\eqref{deltag21} and \eqref{deltag0} for the definition of $\Delta$). Note that only the energy eigenvalues up to 42 lowest eigenstates are plotted in the left panel for the sake of visibility.
}
\end{figure*}
Figure~\ref{fig_sp} shows the low-energy excitation spectrum of a quantum emitter coupled to the cavity array in a broad range of the coupling strength $g$. This spectrum is obtained by the exact diagonalization of the QED Hamiltonian in the AD frame~\eqref{HU1} within the few-photon  ansatz~\eqref{fewphoton} (see Appendix~\ref{app:num} for details about the method). We note that the emitter parameters $v$ and $d$ in Fig.~\ref{fig_sp}  are chosen in such a way that the bare emitter frequency is resonant to the cavity frequency, i.e., $(E_2-E_1)/\hbar\simeq \omega_c$ at $g=0$.

The eigenstates insensitive to $g$ and staying in the photonic band, 
\eqn{
\delta E_{\rm sca}\in[\hbar\omega_{c}-J,\hbar\omega_{c}+J],
} 
correspond to the single-photon scattering states, which are extended over the waveguide and constitute the energy continuum in the thermodynamic limit. 
In contrast, the eigenstates lying out of the band continuum,  
\eqn{
\delta E_{\rm BS}\notin[\hbar\omega_{c}-J,\hbar\omega_{c}+J],
}
behave as the bound states and are accompanied by virtual photons localized around the emitter.  In the nonperturbative regimes, the energies of these bound states  decrease as $\delta E_{\rm BS}\!\propto\!g^{-1}$ and are asymptotically determined by the excitation energies $\delta E_\alpha$ of the decoupled states~\eqref{deceig}. This point is confirmed in the left panel of Fig.~\ref{fig_sp}, where the exact spectrum (blue solid curves) eventually agrees with the asymptotic values (red dashed curves). 

 As discussed earlier, when these bound states lie in the band continuum, 
\eqn{
\delta E_{\rm (q)BIC}\in[\hbar\omega_{c}-J,\hbar\omega_{c}+J],
} 
they behave either as the $\mathbb{Z}_2$-symmetry-protected BIC or as the quasi BIC. These features manifest themselves as the absence of anticrossings in the finite-size spectrum (Fig.~\ref{fig_sp}(a)) or as the presence of tiny  anticrossings with scattering states (Fig.~\ref{fig_sp}(b)), respectively.  We note that this tiny anticrossing of the quasi BIC originates from its vanishingly small decay rate, which can be estimated as (cf. Eq.~\eqref{qbicdec} and the related discussions in Sec.~\ref{subsec:bic})
\eqn{
\Gamma_{\rm qBIC}\sim \frac{(J/\hbar)^2}{g\sqrt{m\omega_c^3d^3}}\left(\frac{v^3}{m_{\rm eff}}\right)^{1/4}\propto g^{-3/2}.
} 
In the ESC regime, the origin of these bound states can also be understood from the fact that the cavity mode spatially overlapping with the emitter is shifted in frequency outside the photonic continuum and thereby hopping to neighboring cavities is strongly suppressed. All the excited light-emitter states within the photonic continuum will then become (quasi) BIC because of this suppression.

We also remark that, in Fig.~\ref{fig_sp}, one can also find several continuum spectra that connect the two-photon continuum with the single-photon one as $g$ is increased. Physically, these states consist of the single-photon scattering states on top of the first, second, and third excited bound states.
The $g$ dependence of those energies can be understood as follows.  They first  rapidly decrease with increasing $g$ until $g/\omega_c\!\sim\!5$. There, bound-state energies  are initially above the height of the potential barrier at $Q\!=\!0$ and hence there are no double degeneracies.  As we increase $g$ further and bound-state energies go well below the potential barrier, these states become nearly degenerate because they now form a pair of symmetric and antisymmetric combinations of excitations localized around each of the two minima in the double-well potential. For instance, in Fig.~\ref{fig_sp} the energy of the third excited state eventually approaches that of the second excited state and they begin to overlap and become doubly degenerate from $g/\omega_c\!>\!5$. Meanwhile, the apparent absence of two-photon scattering states in this regime is motivated by the clarity of presentation, because of which only a certain number of the lowest eigenstates are presented in Fig.~\ref{fig_sp}; this avoids excessive overlaps of the continuous spectra.

\subsection{Dressed potential}
 The effective emitter potential~\eqref{Veff}  in the AD frame is dressed by vacuum electromagnetic fields. In the present case of the double-well potential, the corresponding dressed potential is given by
\eqn{\label{vveff}
V_{{\rm eff}}(Q)=v_{{\rm eff}}\left(1-\frac{Q^{2}}{d_{{\rm eff}}^{2}}\right)^{2},
}
where we neglect an irrelevant constant, introduce the renormalized potential barrier $v_{\rm eff}$ as 
\eqn{\label{veffq}
v_{{\rm eff}}=\begin{cases}
v\left(1-\frac{3\xi^{2}}{d^{2}}\right)^{2} & \xi\leq\frac{d}{\sqrt{3}}\\
0 & \xi>\frac{d}{\sqrt{3}}
\end{cases},
}
and define the effective dipole length by $d_{\rm eff}/d\equiv(v_{\rm eff}/v)^{1/4}$.
As expected from the general argument in Sec.~\ref{subsec:vac}, the barrier $v_{\rm eff}$ is always suppressed compared to the bare value $v$ and the suppression is most significant when $\xi$ becomes maximum, which occurs around the DSC regime (see Fig.~\ref{fig_veff}(a)). Interestingly, when the displacement parameter $\xi$ becomes sufficiently large such that $\xi\!>\!d/\sqrt{3}$, even the full suppression of the potential barrier, i.e., $v_{\rm eff}=0$, is possible. Nevertheless, this does not mean that the entire potential is suppressed because the dipole length $d_{\rm eff}$ in Eq.~\eqref{vveff} also converges to zero in this case. The resulting potential then contains both the quartic and quadratic contributions with the same sign, leading to the single minimum (see Fig.~\ref{fig_veff}(b) for an illustrative example).  
\begin{figure}[t]
\includegraphics[width=86mm]{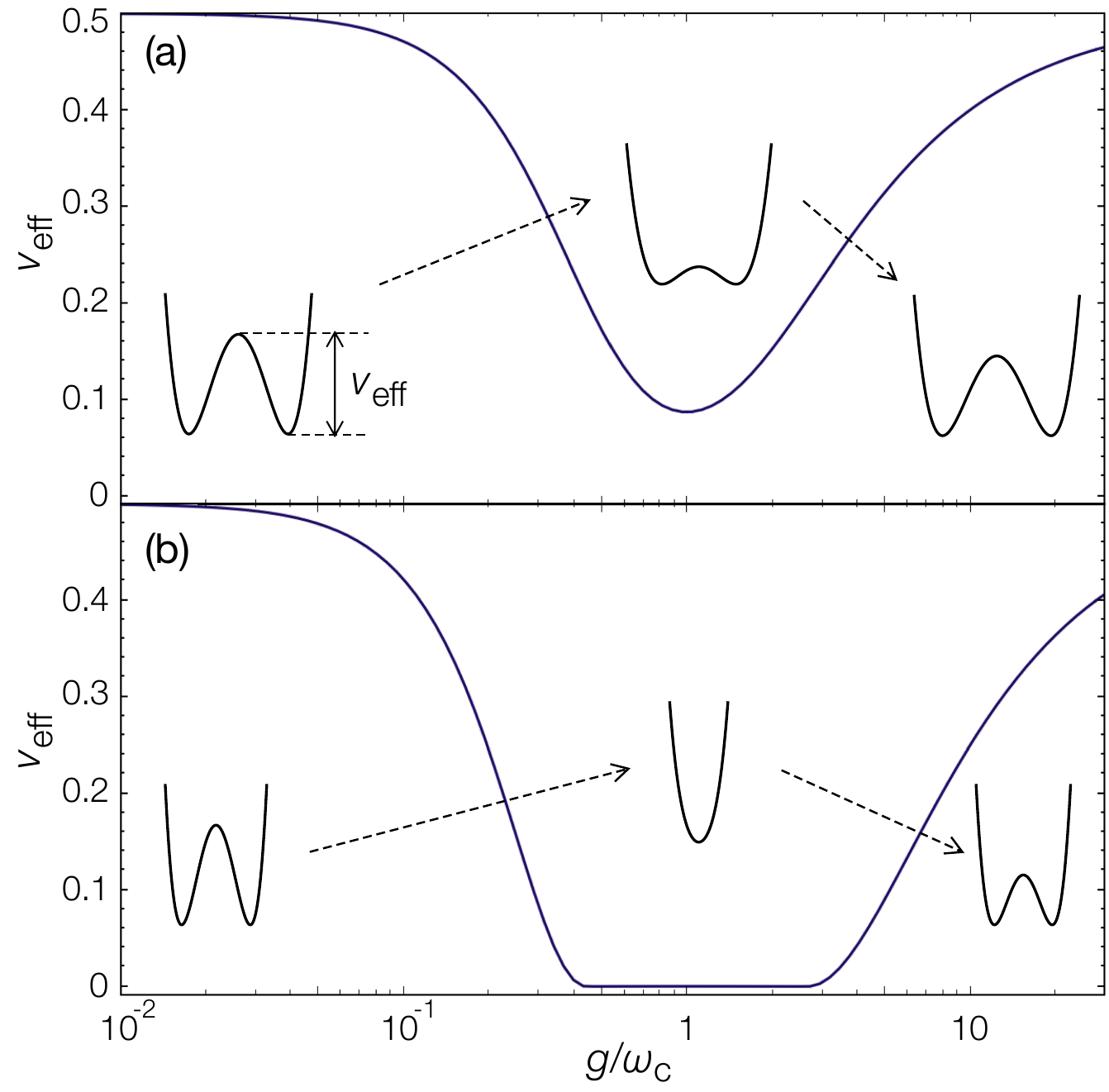} 
\caption{\label{fig_veff}
Potential barrier $v_{\rm eff}$ of the vacuum-dressed effective potential $V_{\rm eff}(Q)$ at different coupling strengths $g$ (see Eq.~\eqref{veffq}). The suppression and eventual restoration of the barrier arises from the nonmonotonic $g$ dependence of $\xi$ in Eq.~\eqref{xiveff} (cf. Fig.~\ref{fig_xi}(b)). Insets show typical spatial profiles of the effective potential in each regime. Parameters are $J=0.1$, $v=0.5$, and $d=1$ in (a), and $J=0.1$, $v=0.5$, and $d=0.6$ in (b).
}
\end{figure}

\subsection{Two-level effective model and its breakdown in the Coulomb gauge\label{subsec:two}}

Construction of the Jaynes-Cummings-type effective model is often useful to simplify the analysis of the original QED Hamiltonian, especially when  low-energy excitations are of interest.
This can be done by performing the two-level truncation of the emitter and assuming the rotating wave approximation.   
In the AD frame, such procedure leads to the standard Jaynes-Cummings  Hamiltonian, but with the suitably {\it renormalized} parameters, 
\eqn{\label{rwaad}
\hat{H}_{U}^{\text{JC}}\!=\!\frac{\hbar\Delta_{g}}{2}\hat{\sigma}^{z}\!+\!\Bigl(\hat{\sigma}^{-}\sum_{n}\hbar\tilde{g}_{n}\hat{b}_{n}^\dagger\!+\!{\rm H.c.}\Bigr)\!+\sum_n\hbar\Omega_n\hat{b}^\dagger_n\hat{b}_n,\nonumber\\
}
where the renormalized emitter frequency $\Delta_g$ and the effective coupling strengths $\tilde{g}_n$ are defined by
\eqn{
\Delta_{g}&\equiv&\frac{E_{2}-E_{1}}{\hbar}>0,\label{deltag21}\\
\tilde{g}_{n}&\equiv&\frac{\xi_{n}}{\hbar}\langle\psi_{1}|\frac{dV}{dQ}|\psi_{2}\rangle.
}
We recall that $E_{1,2}$ ($|\psi_{1,2}\rangle$) represent the two-lowest eigenenergies (eigenstates) of the renormalized emitter Hamiltonian~\eqref{Hmat}, and thus depend on the coupling strength $g$ through $m_{\rm eff}$ and $V_{\rm eff}$.

\begin{figure}[b]
\includegraphics[width=86mm]{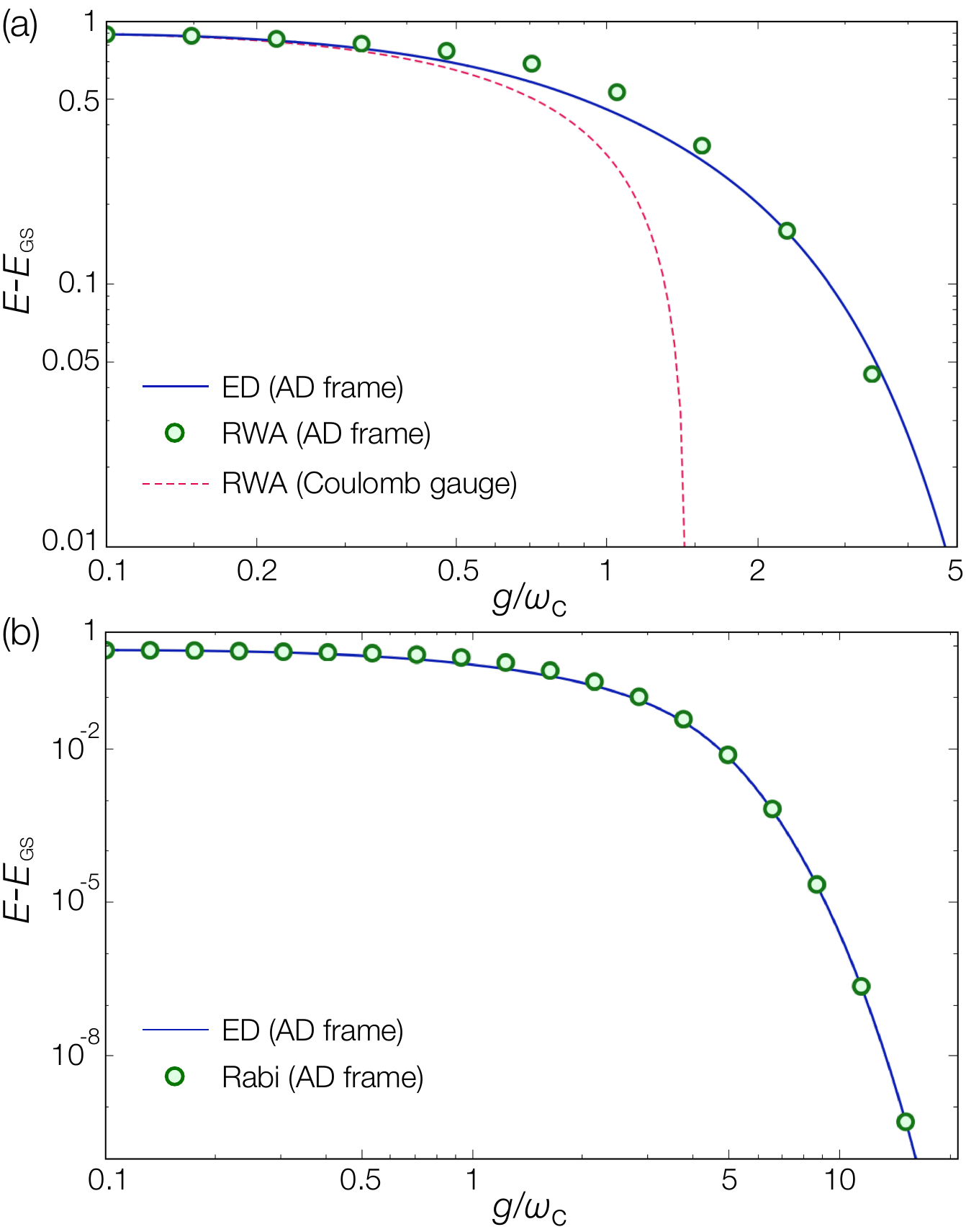} 
\caption{\label{fig_rwa}
Comparisons of the first-excited energy between the exact result and the two-level effective models. The blue solid curves are obtained by the exact diagonalizion (ED) of the AD-frame Hamiltonian~\eqref{HU1}. (a) The green dots represent the results of the two-level model with the rotating wave approximation (RWA) in the AD frame (cf. Eq.~\eqref{rwaad}). The red dashed curve shows the corresponding results in the Coulomb gauge (cf. Eq.~\eqref{rwac}). (b) The green dots represent the results of the quantum Rabi model in the AD frame, i.e., the two-level model without RWA (see Eq.~\eqref{rabi}). 
Parameters are $J=0.1$, $v=0.5$, and $d=0.87$.
}
\end{figure}

Importantly, since the effective spin-bath couplings $\tilde{g}_n$ remain small over the entire region (cf. Fig.~\ref{fig_xi}(b)), the rotating wave approximation in Eq.~\eqref{rwaad} can be performed even beyond the USC regime. Also, the two-level truncation for the emitter in the AD frame here should remain meaningful as discussed in Sec.~\ref{subsec:bre}. We thus expect the effective model $\hat{H}_U^{\text{JC}}$ to be valid not only at weak $g$, but also in the nonperturbative  regimes.  

\begin{figure*}[t]
\includegraphics[width=175mm]{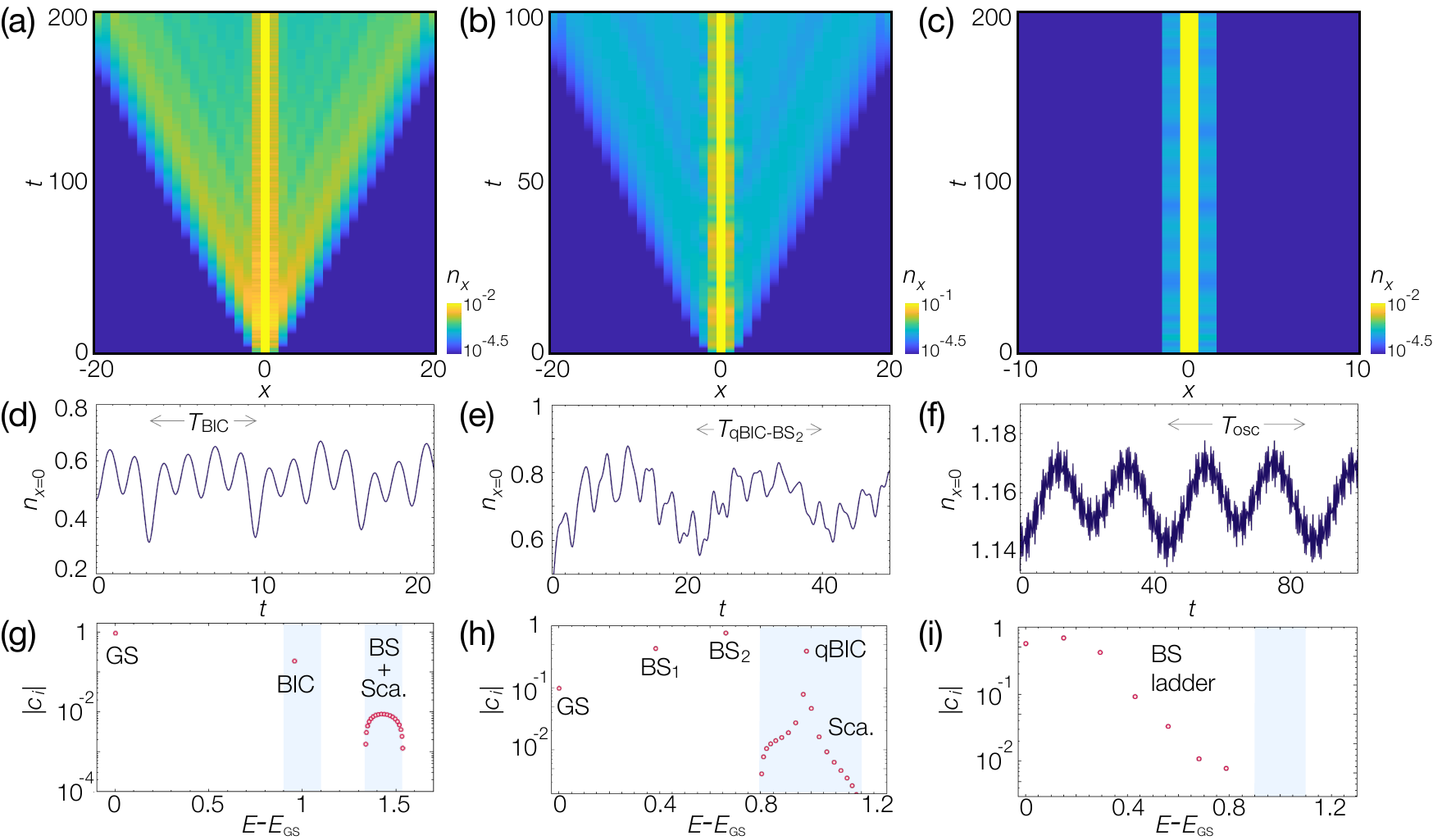} 
\caption{\label{fig_quench}
(a-c) Spatiotemporal dynamics of photon occupancy $n_x=\langle\hat{a}^\dagger_x\hat{a}_x\rangle_{\rm C}$ after the quench at different coupling strengths $g$. (a,b) Excitations of propagating photons in the deep strong coupling regimes, and (c) photon confinement around the emitter at $x\!=\!0$ in the extremely strong coupling regime. 
(d-f) Corresponding dynamics of the photon occupancy $n_{x=0}$  at the emitter position, and (g-i) the initial weights in terms of eigenstates of the post-quench Hamiltonian (cf. Eq.~\eqref{weight}). In (d,e), the excitations of the (quasi) BIC lead to the slow, long-lasting oscillatory dynamics whose period is characterized by the bound-state energies. In (f), a ladder of bound states manifests itself as the oscillatory dynamics with a long period $T_{\rm osc}\!=\!2\pi/\omega_{\rm osc}$ that diverges in the strong-coupling limit (cf. Eq.~\eqref{escosc}). Parameters are 
$g=1.2$, $J=0.1$, $v=0.5$, $d_i=0.6$, and $d_f=0.87$ in (a,d,g),  
  $g=1.3$, $J=0.2$, $v=1$, $d_i=0.9$, and $d_f=2.7$ in (b,e,h), and
 $g=5$,  $J=0.1$, $v=1$, $d_i=2$, and $d_f=2.5$ in (c,f,i).
}
\end{figure*}

To demonstrate this, we plot  in Fig.~\ref{fig_rwa}(a)  the lowest excitation energy $E_{\rm ex}$ of $\hat{H}_U^{\text{JC}}$ which is obtained from 
 the following analytic relation for the single-excitation subspace,
\eqn{\label{bounderg}
E_{{\rm ex}}-\Delta_{g}=\sum_{n}\frac{\tilde{g}_{n}^{2}}{E_{{\rm ex}}-\Omega_{n}}.
}
Indeed, it agrees well with the exact values even in the DSC regime. 
In contrast, the conventional two-level model constructed from the Coulomb-gauge Hamiltonian is given by
\eqn{\label{rwac}
\hat{H}^{\text{JC}}_{{\rm C}}\!=\!\frac{\hbar\Delta}{2}\hat{\sigma}^{z}\!+\!\Bigl(\hat{\sigma}^{-}\sum_{k}\hbar\tilde{g}_{k}\hat{a}_{k}^\dagger\!+\!{\rm H.c.}\Bigr)\!+\!\sum_k\hbar\omega_k\hat{a}^\dagger_k\hat{a}_k,\nonumber\\
}
where $\Delta$ and $\tilde{g}_k$ are the {\it bare} parameters defined by
\eqn{
\Delta&\equiv&\Delta_{g=0},\label{deltag0}\\
\tilde{g}_k&\equiv&\frac{g}{\sqrt{L}}x_{\omega_c}\langle\psi_{1}^{g=0}|\partial_{Q}|\psi_{2}^{g=0}\rangle
}
with $x_{\omega_c}=\sqrt{\hbar/m\omega_c}$. This simplified Hamiltonian $\hat{H}^{\text{JC}}_{{\rm C}}$ takes exactly the same form as $\hat{H}^{\text{JC}}_{{U}}$ in Eq.~\eqref{rwaad}, but with unrenormalized parameters. While this construction is valid at weak $g$, it is well-known that such a na\"{i}ve two-level description in the Coulomb gauge breaks down once one enters into the USC regime in which nonresonant processes become relevant and the two-level truncation becomes ill-justified (see the red dashed curve in Fig.~\ref{fig_rwa}(a)). 

In this respect, the AD frame significantly expands the applicability of the  Jaynes-Cummings description beyond the weak coupling regimes, and thus allows one to use the standard techniques valid within the rotating wave approximation, such as the Wigner-Weisskopf theory, in a broad range of $g$. Nevertheless, we remark that the effective Hamiltonian $\hat{H}^{\text{JC}}_{{U}}$ constructed in the AD frame should also ultimately become invalid when $\Delta_g\!<\!\tilde{g}_n\!\ll\!\omega_c$, where the counter rotating terms turn out to be equally important as rotating ones; this typically occurs in $g\gtrsim 5$. 

Instead, a more accurate description including the counter rotating terms still remains valid even in such ESC regime. Specifically, we can construct the quantum Rabi model in the AD frame, 
\eqn{\label{rabi}
\hat{H}_{U}^{\text{Rabi}}\!=\!\frac{\hbar\Delta_{g}}{2}\hat{\sigma}^{z}\!+\hat{\sigma}^ x\,\!\bigl(\sum_{n}\hbar\tilde{g}_{n}\hat{b}_{n}^\dagger\!+\!{\rm H.c.}\bigr)\!+\sum_n\hbar\Omega_n\hat{b}^\dagger_n\hat{b}_n,\nonumber\\
}
which gives almost the exact results in the ESC regime (see, for instance, the comparison in Fig.~\ref{fig_rwa}(b)). There, we note that the lowest excitation energy exponentially vanishes as $g$ is increased (cf. Eq.~\eqref{renqfreq} below), while the higher excitation energies lie well above this two-level manifold with the energy spacing that scales as $\propto\!1/g$. This is the reason why the quantum Rabi description becomes asymptotically exact in the AD frame.

\subsection{Quench dynamics}

The many-body bound states and the BIC lead to rich nonequilibrium dynamics in the nonperturbative regimes. To be concrete, we consider the quench protocol with the emitter parameter $d$ in the double-well potential~\eqref{doublewell}  being abruptly changed as $d_i\!\to\!d_f$ at time $t\!=\!0$ while keeping all the other parameters constant. This effectively changes the positions of the minima of $V_{\rm eff}$ and also modifies the qubit frequency. 
The initial state $|\Psi(0)\rangle$ is chosen to be the ground state of the QED Hamiltonian at $d\!=\!d_i$ and large fixed $g$. We emphasize that, in the Coulomb gauge, this initial state is already a strongly entangled light-matter state consisting of virtual photons localized around the emitter.   
The quench protocol discussed here should be realized in the current experimental techniques of, e.g., circuit QED that deals with photons in microwave regime. As detailed below, this procedure provides a feasible way to experimentally detect the signature of the predicted many-body BIC, which cannot be excited by a single-photon scattering by definition. 

We calculate the real-time dynamics by transforming to the AD frame, since the analysis in the Coulomb gauge becomes exponentially hard at large $g$ as discussed earlier. Specifically, we exactly diagonalize the post-quench Hamiltonian $\hat{H}_U$ at $d\!=\!d_f$ (see Appendix~\ref{app:num} for details) and obtain the time evolution via  
\eqn{
|\Psi(t)\rangle_{U}&=&e^{-i\hat{H}_{U}t}|\Psi(0)\rangle_{U}\nonumber\\
&=&\sum_{i}c_{i}e^{-iE_{i}t}|\Psi_{i}\rangle_{U},\label{weight}
}
where $E_i$ ($|\Psi_i\rangle_U$) are the corresponding energies (eigenstates), and $c_i$ are expansion coefficients of the initial state. We then calculate the evolution of an observable $\hat{O}$ in the original Coulomb gauge through the unitary transformation 
\eqn{\langle\hat{O}\rangle_{\rm C}\!=\!\langle\hat{U}^\dagger\hat{O}\hat{U}\rangle_U.\label{uctrans}} 
Figure~\ref{fig_quench}(a,b) shows the typical spatiotemporal dynamics of photon occupancy $n_x\!=\!\langle\hat{a}_x^\dagger\hat{a}_x\rangle_{\rm C}$ in the DSC regimes. One can find the nondecaying oscillatory dynamics that is most pronounced around the emitter position $x\!=\!0$ as well as the emission of propagating  photons. The former originates from the existence of the many-body bound states and the (quasi) BIC, while the latter can be considered as the analogue of the dynamical Casimir effect in which physical photons are generated by quenching the vacuum \cite{YE89,SJ92}. 

To gain further insights into the oscillatory dynamics, we plot  the time evolution of the photon occupancy at the origin $x\!=\!0$ in Fig.~\ref{fig_quench}(d,e). We also show the corresponding initial weights $|c_i|$ in Fig.~\ref{fig_quench}(g,h), where the blue shaded regions represent the energy continuum. One can see that the quench protocol excites the (quasi) BIC and the bound states with substantial weights and that their frequencies characterize the long-period oscillation in the dynamics, which typically has a period $T\!=\!{\rm O}(10)$. Thus, those bound states manifest themselves as the long-lasting oscillatory behavior that associates with photons bouncing back and forth around the emitter.

In the ESC regime, photons are so strongly bound by the emitter that  they cannot propagate into the waveguide (see Fig.~\ref{fig_quench}(c)). Besides such photon confinement, the dynamics exhibits the increasingly slow coherent oscillation at larger $g$ (see Fig.~\ref{fig_quench}(f)). This oscillatory behavior  can be understood as follows. In the AD frame, the emitter and photons are asymptotically disentangled and the low-energy dynamics is solely governed by the renormalized emitter Hamiltonian~\eqref{Hmat} with no photon excitations. The present quench protocol then corresponds to the sudden shift of the potential minima of $V_{\rm eff}$. Because the mass is enhanced as $m_{\rm eff}\!\propto\! g^2$ and the wavepacket is tightly localized, this quench initiates the oscillatory dynamics where the wavepacket (initially localized at $d\!\simeq \!d_i$) oscillates around the new minima at $d\!\simeq\! d_f$. Such oscillation frequency can be estimated as
\eqn{\label{escosc}
\omega_{{\rm osc}}=\sqrt{\frac{8v}{d_{f}^{2}m_{{\rm eff}}}\left(1-\frac{3\xi^{2}}{d_{f}^{2}}\right)}\propto g^{-1},
}
which vanishes in the strong-coupling limit. The estimated period $T_{\rm osc}\!=\!2\pi/\omega_{\rm osc}$ agrees well with the observed long-period oscillation in Fig.~\ref{fig_quench}(f). 
In the energy basis, this slow coherent  dynamics manifests itself as excitations of a ladder of bound states corresponding to the decoupled eigenstates~\eqref{deceig} (see Fig.~\ref{fig_quench}(i)). 

\section{Extension to multiple quantum emitters \label{sec:cas}}

We now extend our theoretical formalism to the case of multiple quantum emitters. We discuss several limiting cases and construct the effective two-level model that is valid in a broad range of the light-matter coupling strength.

\subsection{General formalism}
We consider $N$ emitters of mass $m_{j}$ that are subject to potential $V_j$ and locally interact with the common electromagnetic modes at positions $x_{j}$ with $j\!=\!1,2,\ldots,N$. We thus start from the multi-emitter QED Hamiltonian in the Coulomb gauge, 
\eqn{
\hat{H}_{{\rm C}}\!=\!\!\sum_{j=1}^{N}\Biggl[\frac{\left(\hat{P}_{j}\!-\!q\hat{A}_{x_{j}}\right)^{2}}{2m_{j}}\!+\!V_j(\hat{Q}_{j})\Biggr]\!+\!\sum_{k}\hbar\omega_{k}\hat{a}_{k}^{\dagger}\hat{a}_{k},
}
where the position and momentum operators of the emitters satisfy
\eqn{
[\hat{Q}_{i},\hat{P}_{j}]=i\hbar\delta_{ij},
}
and the vector potential is given by
\eqn{
\hat{A}_{x_{j}}=\sum_{k}f_{kj}\left(\hat{a}_{k}e^{ikx_{j}}+\hat{a}_{k}^{\dagger}e^{-ikx_{j}}\right)
}
with $f_{kj}$ characterizing  an electromagnetic coupling between photonic mode $k$ and emitter $j$; for the sake of simplicity, we assume $f_{kj}=f_{-kj}$. 

Generalizing the unitary transformation~\eqref{unitary} to such multi-emitter case, we obtain the following Hamiltonian (see Appendix~\ref{app:multi} for details):
\eqn{\label{multihu}
\hat{H}_{U}=\hat{H}_{{\rm matter}}+\hat{H}_{{\rm int}}+\hat{H}_{{\rm light}},
}
where the emitter part is given by
\eqn{\label{multimat}
\hat{H}_{{\rm matter}}\!=\!\sum_{j}\Bigl[\frac{\hat{P}_{j}^{2}}{2m_{{\rm eff},j}}\!+\!V_{{\rm eff},j}(\hat{Q}_{j})\Bigr]\!+\!\!\sum_{i>j}\mu_{ij}\hat{P}_{i}\hat{P}_{j}
}
with 
\eqn{
m_{{\rm eff},j}&=&m_{j}/\left[(1+2G)^{-1}\right]_{jj},\label{multimeff}\\
\mu_{ij}&=&\left[(1+2G)^{-1}\right]_{ij}/\sqrt{m_{i}m_{j}}.\label{multimu}
}
Here, $G$ is the $N\!\times\!N$ matrix defined by
\eqn{G_{ij}&\equiv&\sum_{k}\frac{g_{ki}g_{kj}}{\omega_{k}^{2}}\cos(k(x_{i}-x_{j})),\\
g_{kj}&\equiv&qf_{kj}\sqrt{\frac{\omega_{k}}{m_{j}\hbar}}.
}
Physically, $m_{{\rm eff},j}$ in Eq.~\eqref{multimeff} represents the renormalized mass similar to $m_{\rm eff}$ in the single-emitter case considered before except for its dependence on emitter positions through $G$. The coupling $\mu_{ij}$ in Eq.~\eqref{multimu} represents the waveguide-mediated interaction between emitters; in the original Coulomb gauge, this corresponds to the nondissipative coupling mediated by virtual photons in the waveguide.  

In the renormalized multi-emitter Hamiltonian~\eqref{multimat}, the vacuum-dressed effective potentials $V_{{\rm eff},j}$ are given by
\eqn{
V_{{\rm eff},j}(Q)&\equiv &V_j(Q)+\sum_{l=1}\frac{\xi_{j}^{2l}}{(2l)!!}V_j^{(2l)}(Q),\\
\xi_{j}^{2}&\equiv&\sum_{n}|\xi_{nj}|^{2}.
}
The interaction Hamiltonian is 
\eqn{
\hat{H}_{{\rm int}}=\sum_{j}:V_j(\hat{Q}_{j}+\hat{\Xi}_{j}):
}
with
\eqn{
\hat{\Xi}_{j}\equiv\sum_{n}i(\xi_{nj}\hat{b}_{n}^{\dagger}-\xi_{nj}^{*}\hat{b}_{n}),
}
where the displacement parameters $\xi_{nj}$ characterize the effective coupling strengths between dressed photon mode $n$ and emitter $j$ (see Appendix~\ref{app:multi}). The photon part $\hat{H}_{\rm light}$ takes the same form as the single-emitter case in Eq.~\eqref{photonham} \footnote{We emphasize that dressed photon modes discussed here are in general distinct from those in the single-emitter system, as inferred from the sensitivity to emitter positions in multi-emitter cases (see Appendix~\ref{app:multi}).  Nevertheless, we shall use the same subscript $n$ to label electromagnetic mode for the sake of notational simplicity.}. 

To be concrete, from now on we consider the case of identical emitters with 
$m_j=m$, $V_j=V$, and $f_{kj}=f_{k}\;\forall j$. This simplification leads to the identical bare light-matter couplings, $g_{kj}=g_{k}\;\forall j$. In contrast, we note that the effective masses $m_{{\rm eff},j}$, the displacement parameters $\xi_{nj}$, and the dressed potentials  $V_{{\rm eff},j}$ can still depend on the emitter positions, and thus we need subscript $j$ to distinguish them in general. Below we illustrate aspects of several limiting cases, but leave the full understanding of multi-emitter waveguide QED systems to a future work.

\subsection{Two emitters}
\begin{figure}[b]
\includegraphics[width=86mm]{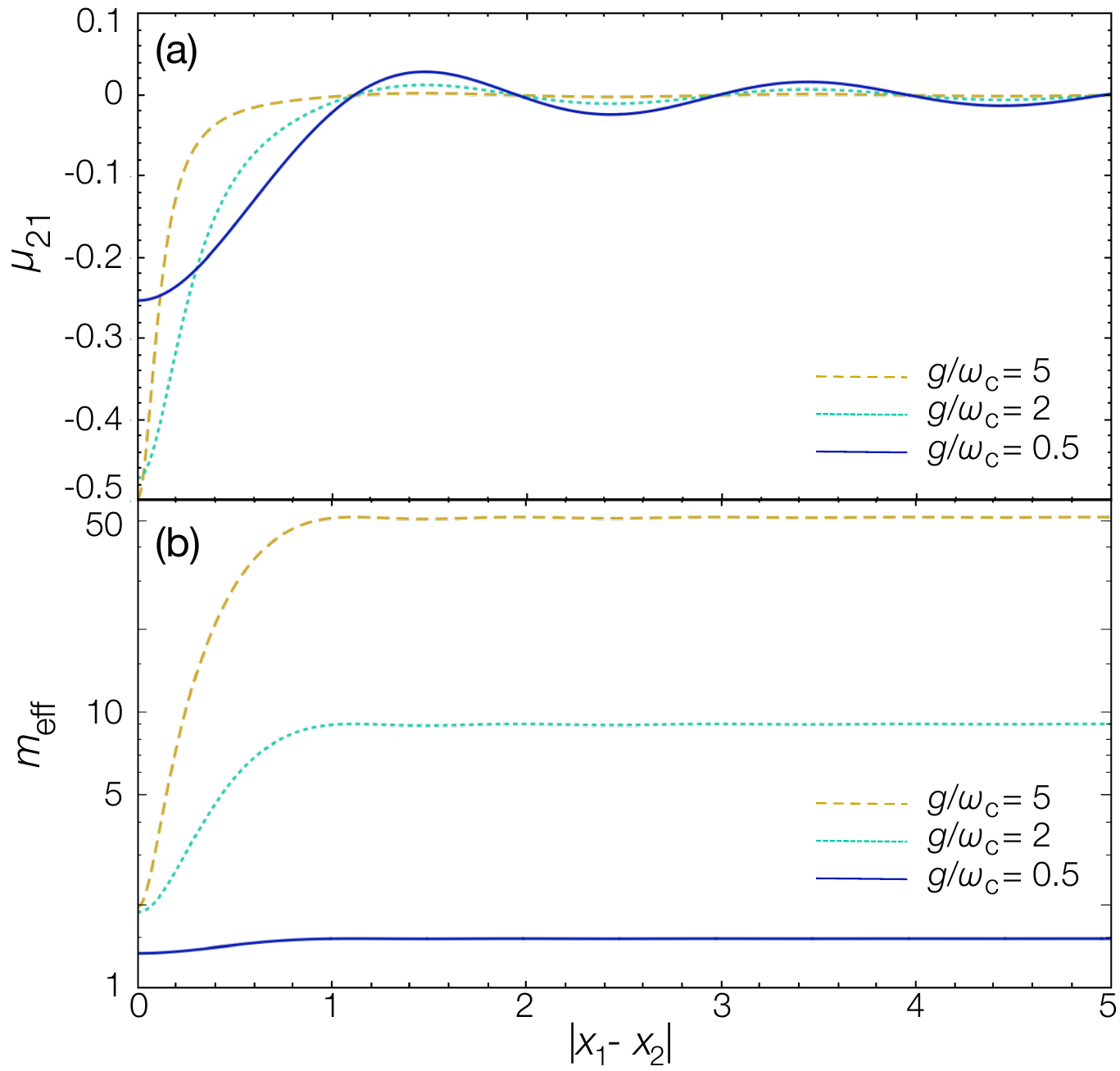} 
\caption{\label{fig_G12}
Renormalized parameters in the two-emitter case. (a) Waveguide-mediated emitter-emitter interaction strength $\mu_{21}$ in Eq.~\eqref{multimu} and (b) effective mass $m_{\rm eff}$ in Eq.~\eqref{multimeff} are plotted against the emitter separation at different light-matter couplings $g$. The waveguide is assumed to be the coupled cavity array as in Sec.~\ref{sec:app} and we set $J=0.2$.
}
\end{figure}
We begin by discussing the two-emitter case. Figure~\ref{fig_G12} plots the waveguide-mediated coupling $\mu_{21}$ and the effective mass $m_{\rm eff}$ against the emitter separation at different coupling strengths $g$. We here assume the waveguide to be the same cavity array as considered in Sec.~\ref{sec:app}. In the USC regime, the coupling $\mu_{21}$ exhibits the oscillatory behavior and can be long-ranged. As $g$ is further increased,  it becomes increasingly short-ranged with oscillations being damped (see Fig.~\ref{fig_G12}(a)).  Such suppression can be interpreted as a nonperturbative effect originating from the tighter confinement of virtual photons around the emitters. 

Figure~\ref{fig_G12}(b) shows that the effective mass monotonically increases at larger $g$. As noted earlier, the effective mass in multi-emitter systems is sensitive to the emitter separation; for the two-emitter case, it starts from $m\frac{1+4\Theta}{1+2\Theta}$ at zero separation and eventually saturates to the single-emitter limit $m(1\!+\!2\Theta)$ when the separation surpasses the cavity length $x_{\omega_c}\!=\!1$ (cf. Eq.~\eqref{meff} and \eqref{mefftheta}).

\subsection{Localized $N$ emitters\label{subsec:loc}}
We next consider the case in which all the emitters are coupled to the waveguide at the same position $x_{1}\!=\!\cdots\!=\!x_{N}\!=\!0$. This corresponds to the case of $N$ (artificial) atoms collectively coupled to the common electromagnetic fields, which is relevant to various experimental setups. It is useful to introduce the collective   momentum and coordinate  by 
\eqn{
\hat{P}_{{\rm CM}}&\equiv&\frac{1}{\sqrt{N}}\sum_{j}\hat{P}_{j},\;\;\;\;\hat{Q}_{{\rm CM}}\equiv\frac{1}{\sqrt{N}}\sum_{j}\hat{Q}_{j},
}
as well as the relative ones via
\eqn{
\hat{p}_{j}&\equiv&\hat{P}_{j}-\frac{\hat{P}_{{\rm CM}}}{\sqrt{N}},\;\;\;\;\hat{q}_{j}\equiv\hat{Q}_{j}-\frac{\hat{Q}_{{\rm CM}}}{\sqrt{N}}.
}
The Hamiltonian can be rewritten as
\eqn{\label{collecham}
\hat{H}_{U}=\hat{H}_{{\rm CM}}+\hat{H}_{{\rm int}}+\hat{H}_{{\rm light}}+\hat{H}_{{\rm rel}},\label{com}
}
where
\eqn{\label{collechammat}
\hat{H}_{{\rm CM}}=\frac{\hat{P}_{{\rm CM}}^{2}}{2M_{{\rm eff}}}+N\,V_{{\rm eff}}\Bigl(\frac{\hat{Q}_{{\rm CM}}}{\sqrt{N}}\Bigr)
}
governs the dynamics of the collective mode with the effective mass 
\eqn{\label{meffmulti}
M_{{\rm eff}}=m\left(1+2N\Theta\right).
}
The electromagnetic interaction between the collective mode and photons is given by 
\eqn{
\hat{H}_{{\rm int}}=\; :N\,V\Bigl(\frac{\hat{Q}_{{\rm CM}}}{\sqrt{N}}+\hat{\Xi}\Bigr):.
}
The relative motion of emitters is governed by 
\eqn{
\hat{H}_{{\rm rel}}\!=\!\sum_{j}\Bigl[\frac{\hat{p}_{j}^{2}}{2m}\!+\!V\Bigl(\hat{q}_{j}\!+\!\frac{\hat{Q}_{{\rm CM}}}{\sqrt{N}}\!+\!\hat{\Xi}\Bigr)\!-\!V\Bigl(\frac{\hat{Q}_{{\rm CM}}}{\sqrt{N}}\!+\!\hat{\Xi}\Bigr)\Bigr].\nonumber\\
}
Here we recall that the relative variables satisfy the constraints $\sum_j\hat{p}_j\!=\!\sum_j\hat{q}_j\!=\!0$ and thus they contain $N\!-\!1$ degrees of freedom. 
When the collective mode dominantly couples to the electromagnetic fields,  one may neglect fluctuations and dynamics of the relative degrees of freedom.  The total QED  Hamiltonian~\eqref{collecham} then becomes equivalent to the single-emitter one~\eqref{HU1} upon  the replacements $\hat{Q}\!\to\!\hat{Q}_{{\rm CM}},$ $\hat{P}\!\to\!\hat{P}_{{\rm CM}}$, $g_{k}\!\to\!\sqrt{N}g_{k}$, $d\!\to\!\sqrt{N}d$, and $v\!\to\! Nv$. While this equivalence is nothing but the well-known $\sqrt{N}$ collective enhancement of the dipole and the coupling strength, our analysis indicates that it can remain even in the DSC/ESC regimes where the multilevel nature of emitters becomes crucial, as long as relative motion does not play a substantial role. 

\subsection{Two-level effective model}
We next extend the construction of the two-level effective model discussed in Sec.~\ref{subsec:two} to arbitrary multi-emitter systems.  The projection onto the two-lowest dressed emitter states in the AD frame results in the effective model 
\eqn{\label{multieff}
\hat{H}_U=\hat{{H}}_{\text{Ising}}+\hat{H}^{\text{JC}}_{\rm int}+\hat{H}_{\rm light},
}
where the matter part corresponds to the (inhomogeneous) transverse-field Ising model,
\eqn{\label{ising}
\hat{{H}}_{\text{Ising}}=\sum_{j}\frac{\hbar\Delta_{g,j}}{2}\,\hat{\sigma}_{j}^{z}+\sum_{i>j}J_{ij}\,\hat{\sigma}_{i}^{x}\hat{\sigma}_{j}^{x}
}
with $J_{ij}$ being the waveguide-mediated qubit-qubit interaction given by
\eqn{
J_{ij}=-\hbar^2\mu_{ij}\langle\psi_{1i}|{\partial}_{Q_i}|\psi_{2i}\rangle\langle\psi_{1j}|{\partial}_{Q_j}|\psi_{2j}\rangle.
}
Here, $|\psi_{1,2j}\rangle$ represent the two-lowest eigenstates of the renormalized emitter Hamiltonian $\hat{P}_{j}^{2}/2m_{{\rm eff},j}\!\!+\!\!V_{{\rm eff},j}$, and $\Delta_{g,j}\!\geq\!0$ is the corresponding excitation energy.
We recall that $\Delta_{g,j}$ depends on emitter positions through $m_{{\rm eff},j}$ and $V_{{\rm eff},j}$.
The Jaynes-Cummings-type light-matter interaction is 
\eqn{\label{JCising}
\hat{H}^{\rm JC}_{\rm int}=\sum_j\hat{\sigma}^{-}_j\sum_{n}\hbar\tilde{g}_{nj}\hat{b}_{n}^\dagger\!+\!{\rm H.c.},
}
where $\tilde{g}_{nj}$ are the effective qubit-boson couplings given by
\eqn{
\tilde{g}_{nj}=\frac{\xi_{nj}}{\hbar}\langle\psi_{1j}|\frac{dV_j}{dQ}|\psi_{2j}\rangle.
}
 As discussed before, in contrast to the Coulomb/PZW gauges, our construction in the AD frame should remain valid even at large $g$ because the level truncations can increasingly be well-justified in the strong-coupling limit. Thus, the two-level effective model~\eqref{multieff} can be used to accurately capture low-energy physics of the original multi-emitter QED Hamiltonian  in a broad range of the coupling strength. Nevertheless, we note that in the ESC regime the counterrotating terms can be important and, in such a case, the Rabi-type interaction instead of Eq.~\eqref{JCising} should give more accurate results.
In the single-emitter case, we recall that the asymptotic decoupling and the enhanced photon frequency led to  the decoupled eigenstates~\eqref{deceig}.  
Similarly, in the present multi-emitter case, one may set the photon state to be the vacuum and reduce the whole problem to the Ising Hamiltonian~\eqref{ising}, from which the ground-state properties and elementary excitations can be extracted. To make the results quantitatively accurate, one can extend the analysis to the few-photon sector in the same manner as done in Sec.~\ref{sec:app} when necessary.

It is worthwhile to discuss a simple case of homogeneous configuration, in which the emitters are periodically placed with the same separation. In this case, the renormalized frequency and the spin-boson couplings are independent of an emitter, $\Delta_{g,j}=\Delta_{g}$, $\tilde{g}_{nj}=\tilde{g}_n$ $\forall j$, and the spin-spin interaction becomes translationally invariant $J_{ij}=J_{|i-j|}$. 
Then, the multi-emitter Hamiltonian~\eqref{ising} reduces to the standard homogeneous transverse-field Ising model with couplings $J_{|i-j|}$ that are in general long-ranged.
In particular, in the limit of zero emitter separation, the effective Hamiltonian reduces to the Lipkin-Meshkov-Glick (LMG) model:
\eqn{
\hat{H}_{\rm LMG}=\frac{\hbar\Delta_g}{2}\hat{S}^z+J'(\hat{S}^x)^2,
}
where $J'\!>\!0$ is the all-to-all antiferromagnetic coupling, $\hat{S}^{\gamma}\!\equiv\!\sum_j\hat{\sigma}_j^\gamma$ are the collective spin operators with $\gamma\!\in\!\{x,y,z\}$, and we neglect the irrelevant constant.

These simplifications in the AD frame allow us to export the insights and techniques originally developed in studies of the transverse-field Ising models to the analysis of the multi-emitter QED Hamiltonian in the nonperturbative regimes. Indeed, it is known that, in the many-emitter limit $N\!\to\!\infty$, such model exhibits rich phase diagrams depending on dimension, lattice geometry, or sign/decay exponent of the long-range coupling $J_{|i-j|}$ \cite{SS12,HS16,SSN18,FS19}. Moreover, a disordered transverse-field Ising model is argued to realize many-body localization \cite{KJA14,HP15}, and such disorder is fairly ubiquitous in the multi-emitter Hamiltonian~\eqref{ising} where disorder comes into play through emitter positions. 
While we leave the full understanding of such multi-emitter physics at strong light-matter couplings to future investigations, our analysis provides a reliable starting point for this and shows promise for realizing the above exotic phases in the waveguide QED.    

\section{Quantum phase transitions with gapless dispersions \label{sec:gless}}
We finally turn our attention to the case of gapless dispersions. Specifically, we consider the photon frequencies  that, in the low-energy limit, scale as
\eqn{
\omega_{k}\propto k^{l},
}
where $l\!>\!0$ is an exponent characterizing the gapless dispersion. 
This type of dispersions can be realized in waveguide QED systems by using transmission lines or by designing mode frequencies with fabricated resonators.
 One of the key questions here is whether or not a waveguide QED system governed by the Hamiltonian~\eqref{HC} undergoes a quantum phase transition as the light-matter coupling is increased. Below we discuss that the presence or absence of transition can be understood in terms of the mass renormalization after the unitary transformation, and demonstrate it by analyzing a concrete model of circuit QED.

\subsection{Delocalization-localization transition and mass renormalization}
The ground state of a single-emitter system displays either delocalized or localized phase that is characterized by the following order parameter
\eqn{
{\cal O}&\equiv&\lim_{h\to+0}\lim_{L\to\infty}\langle\hat{Q}\rangle_{h,{\rm C}}\\
&=&\lim_{h\to+0}\lim_{L\to\infty}\langle\hat{Q}+\hat{\Xi}\rangle_{h,U},\label{orderU}
}
where $\langle\hat{Q}\rangle_{h,{{\rm C}}}$ represents an emitter displacement in the ground state of the QED Hamiltonian in the Coulomb gauge~\eqref{HC} with a bias potential $-h\hat{Q}$ being added to $V(\hat{Q})$; from now on, we  assume $V$ to be the standard double-well potential~\eqref{doublewell} though our arguments can be applied to a generic potential profile. In the AD frame, this order parameter corresponds to an expectation value of $\langle\hat{Q}\!+\!\hat{\Xi}\rangle_{U}$ (cf. Eq.~\eqref{uctrans}).
The delocalized phase is characterized by the vanishing order parameter ${\cal O}\!=\!0$ and the unique, nondegenerate ground state. In contrast,  in the localized phase, the ground state exhibits the two-fold degeneracy in the thermodynamic limit $L\!\to\!\infty$  corresponding to localization to each of the two minima in the potential. In this case, the order parameter takes a nonzero value ${\cal O}\!>\!0$, which indicates the broken $\mathbb{Z}_2$ symmetry in Eq.~\eqref{z2sym}. 


At sufficiently large coupling $g$, the emitter and photons are decoupled  in the AD frame, and the first excitation energy $\Delta_g$ can be estimated from the tunneling rate in the dressed potential~\eqref{Veff}, resulting in \cite{JLG81}
\eqn{\label{renqfreq}
\hbar\Delta_g\simeq \frac{\hbar^2}{m_{\rm eff}d_{\rm eff}^2}\exp\left[-\frac{4}{3}\sqrt{\frac{2m_{\rm eff}d_{\rm eff}^2v_{\rm eff}}{\hbar^2}}\right].
}
Thus, the divergent $m_{\rm eff}$ leads to the gap closing $\lim_{L\to\infty}\Delta_g\!=\!0$, indicating a possible ground-state degeneracy, i.e., transition to the localized phase. In contrast, as long as $m_{\rm eff}$ remains finite, such exact two-fold degeneracy of the ground state is unlikely to happen; this implies the absence of transition. We delineate general properties in each of these cases on the basis of this observation.

Firstly, when the effective mass remains finite $m_{\rm eff}\!<\!\infty$ the whole results discussed in Secs.~\ref{sec:lm}-\ref{sec:app} for a gapped dispersion qualitatively remain the same, except for the point that all the bound states now behave as the (quasi) BIC in the present gapless case. Importantly, the ground state can thus be well-approximated by the lowest decoupled eigenstate $|\psi_{0}\rangle|0\rangle$ (cf. Eq.~\eqref{deceig}), which has ${\cal O}\!=\!0$ and provides the unique ground state due to the nonvanishing excitation energy $\Delta_g\!>\!0$ (see Eq.~\eqref{renqfreq}). Note that, while $\Delta_g$ can be exponentially small as $g$ is increased, it still remains nonzero in $L\!\to\!\infty$ at any finite $g$. Thus, the ground state is not expected to exhibit the exact two-fold degeneracy and should remain delocalized. It is worthwhile to note that the same argument should also rule out the possibility of a phase transition for general gapped dispersions, which include some experimentally relevant situations, such as cavity array and (finite-size) open transmission lines. 

Secondly, in certain gapless dispersions, the effective mass $m_{\rm eff}$ exhibits the infrared divergence and grows polynomially as a function of system size $L$. One can also check that this leads to the polynomial divergence of $\xi$ in the dressed emitter potential~\eqref{Veff}; the resulting effective potential $V_{\rm eff}$ then has the unique minimum at $Q\!=\!0$ (cf. Eq.~\eqref{veffq}) and  becomes infinitely tight in $L\!\to\!\infty$. Thus, in the thermodynamic limit, the emitter wavefunction in the AD frame is completely localized at $Q\!=\!0$, and the total system is solely governed by  the photonic part
\eqn{\label{subohmic}
\hat{H}_{U}=\; :V(\hat{\Xi}):+\sum_{n}\hbar\Omega_{n}\hat{b}_{n}^{\dagger}\hat{b}_{n}.
}
An order of its ground state is characterized by the expectation value of $\langle\hat{\Xi}\rangle_{U}$ (see Eq.~\eqref{orderU}). 
In the limit of a deep potential with large $v$, the first term in Eq.~\eqref{subohmic} is dominant, and the ground state should exhibit the two-fold degeneracy at $\langle\hat{\Xi}\rangle_U\!=\!\pm d$ in $L\!\to\!\infty$. This leads to the localized phase with ${\cal O}>0$. In contrast, in the opposite limit of a shallow potential, the second (harmonic) term in Eq.~\eqref{subohmic} dominates over the potential term, and the vacuum state gives the unique ground state, which has $\langle\hat{\Xi}\rangle_U\!=\!0$ and leads to the delocalized phase with ${\cal O}\!=\!0$. Finally, in between these two limits, the potential landscape effectively changes from the double-well profile to the harmonic one, and accordingly,  the order parameter $\cal O$ continuously decreases from $d$ and vanishes at  certain $v^*\!>\!0$. 
We recall that, in general, a deep (shallow) potential depth $v$ corresponds to a small (large) bare qubit frequency $\Delta$. 
To sum up, one expects a continuous quantum phase transition between the localized phase at strong $v$ (resp. small $\Delta$) and the delocalized phase at weak $v$ (resp. large $\Delta$); see the blue dashed vertical arrow in Fig.~\ref{fig_phase}.

\begin{figure}[t]
\includegraphics[width=80mm]{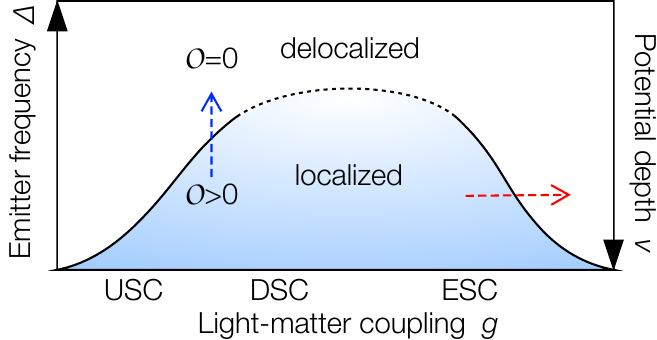} 
\caption{\label{fig_phase}
Schematic figure illustrating the ground-state phase diagram of the QED Hamiltonian~\eqref{HC}. A salient feature which is not present in the simplified spin-boson models is that the waveguide QED system considered here should ultimately reenter into the delocalized phase at a sufficiently large light-matter coupling (red dashed horizontal arrow).  
}
\end{figure}

Finally, the present consideration of the full QED Hamiltonian in nonperturbative regimes also reveals an intriguing new possibility beyond what is commonly expected before. Namely, as inferred from the nonmonotonic $g$ dependence of the coupling coefficients $\zeta_n$ in Eq.~\eqref{HCb} and accordingly $\xi_n$ (cf. Fig.~\ref{fig_xi}(b)), the system should again transition into the delocalized phase at a sufficiently large light-matter coupling (see the red dashed horizontal arrow in Fig.~\ref{fig_phase}).  This is expected to occur in the ESC regime, where the coupling strength dominates all the other energy scales. 
Physically, the origin of this favoring of the delocalized phase can be traced back to the diamagnetic $\hat{A}^2$ term that suppresses the displacements of the bosonic modes from the vacuum. The latter prohibits populating a macroscopic number of low-momentum photons, which is necessary to induce the transition to the localized phase \cite{SH85}; this is reminiscent of what has been discussed in the context of the no-go theorems of the superradiant transition \cite{RK75,NP10,VO11,DB182,AGM19,SA19}.

\subsection{Case study of the circuit QED Hamiltonian\label{sec:frg}}

One may expect that the present QED Hamiltonian with the double-well potential should feature the similar physics as known for the spin-boson model. For instance, it is often supposed that the two-level projection of the PZW Hamiltonian~\eqref{Hpzw} should allow for the spin-boson description of the waveguide QED systems. However, as discussed before, such level-truncation procedure cannot in general be justified at strong couplings, and we must carefully reexamine the ground-state properties of nonperturbative waveguide QED systems separately from the simplified spin-boson description. Indeed, in the previous section, we point out that the full-fledged QED systems should exhibit a new feature, which is not present in the usual spin-boson models \cite{BR03}, such as the {\it reentrant transition} into the delocalized phase at sufficiently large light-matter coupling. 

In this section, we concretely demonstrate these results in the case of  resistively shunted Josephson junctions by using the functional renormalization group (FRG) analysis. For the sake of convenience, we here switch to the notation familiar with the circuit QED community. Specifically, we consider a microscopic circuit Hamiltonian (we set $\hbar\!=\!1$) \cite{KM21}:
\eqn{\label{Hcir}
\hat{H}_{{\rm C}}\!=\!E_{C}\hat{N}^{2}\!+\!V(\varphi)\!-\!\hat{N}\sum_{m=1}^{M-1}\zeta_{m}(\hat{b}_{m}\!+\!\hat{b}_{m}^{\dagger})\!+\!\sum_{m=1}^{M-1}\Omega_{m}\hat{b}_{m}^{\dagger}\hat{b}_{m},\nonumber\\
}
where $\varphi$ is the Josephson junction phase and $\hat{N}\!=\!-i\partial/\partial\varphi$ is the charge operator, which play roles as the position $\hat{Q}$ and momentum $\hat{P}$ operators in the previous notation, respectively. The form of the Hamiltonian~\eqref{Hcir} precisely corresponds to the Coulomb-gauge-type Hamiltonian in Eq.~\eqref{HCb} obtained after the Bogoliubov transformation. The charging energy is denoted by $E_{\rm C}$ and the potential term $V(\varphi)$ is chosen to be the double-well potential in the same way as before: 
\eqn{\label{dwdef}
V(\varphi)=\frac{v}{8}\left(\varphi^2-\varphi^2_0\right)^2.
}
In practice, such potential is routinely realized in flux qubits by combining the inductive energy and flux-tuned Josephson energy.  
The coupling coefficient $\zeta_m$ and the environmental frequency $\Omega_m$ of mode $m\in\{1,2,\ldots,M-1\}$ are given by
\eqn{\label{zetacp}
\zeta_{m}&=&\sqrt{\frac{\pi W\Omega_{m}}{M\alpha\left(1+\left[\left(\frac{\pi W}{\alpha E_{C}}-1\right)\tan\left(\frac{m\pi}{2M}\right)\right]^{2}\right)}},\\
\Omega_{m}&=&W\sin\left(\frac{m\pi}{2M}\right),
} 
where $W$ is the environmental cutoff frequency and $\alpha$ is the dimensionless parameter characterizing the coupling strength; the latter is related to the shunt resistance $R$ via $\alpha\!=\!R_Q/R$ with $R_Q\!=\!h/(4e^2)$ being the quantum of resistance. The characteristic coupling strength $g$ and environmental frequency $\omega$ defined in Eqs.~\eqref{gdef} and \eqref{pome} basically correspond to $\sqrt{\alpha E_{\rm C}W}$ and $W$ in the present notation, respectively. Thus, for instance, the ESC regime corresponds to the region $\alpha E_{\rm C}\gtrsim W$.  
We note that, when the environmental cutoff $W$ satisfies the weak coupling condition $\alpha E_{\rm C}\!\ll\! W$, the low-energy limit ($m\!\ll\! M$) of Eq.~\eqref{Hcir} reproduces the Caldeira-Leggett description of the Ohmic dissipation \cite{KM21,LAJ87,Affleck01}. Since we are here interested in the ESC regime $\alpha E_{\rm C}\gtrsim W$, we need to analyze the original circuit QED Hamiltonian~\eqref{Hcir} without taking such limit.

\begin{figure}[b]
\includegraphics[width=80mm]{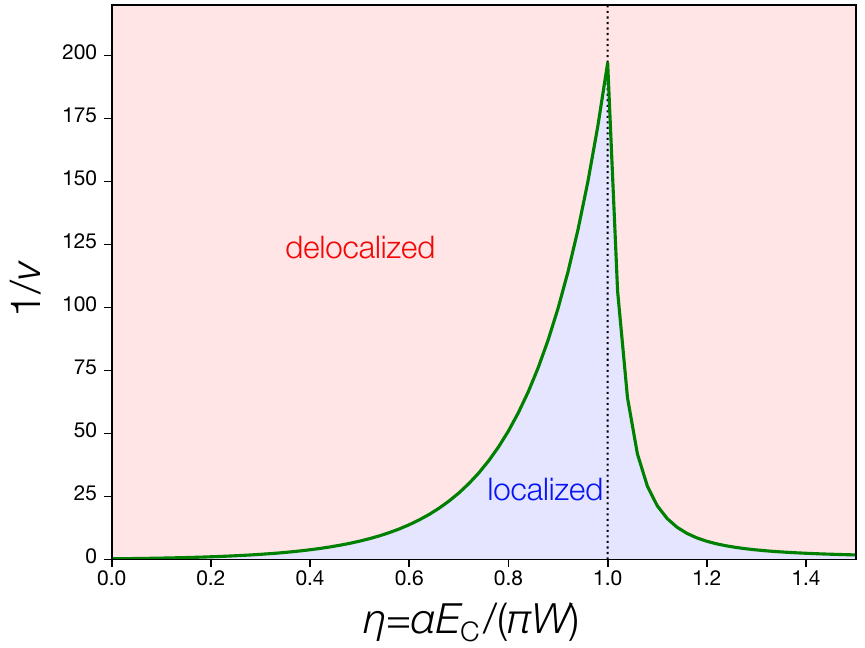} 
\caption{\label{fig_frg}
Ground-state phase diagram of the circuit QED Hamiltonian~\eqref{Hcir} determined from the FRG analysis. The vertical (horizontal) axis represents the inverse of the potential barrier (the normalized coupling strength); the potential barrier $v$ is defined in Eq.~\eqref{dwdef}.  A reentrant transition to the delocalized phase can occur in the ESC regime where the coupling strength $\alpha E_{\rm C}$ dominates over the other energy scales including environmental cutoff $W$ (compare it with Fig.~\ref{fig_phase}). We set the FRG UV cutoff $\Lambda_0\!=\!1$ as the energy unit, and choose the parameters $E_{\rm C}\!=\!1$, $W\!=\!30$, and $\varphi_0\!=\!1$.
}
\end{figure}

Before providing a quantitative analysis, we illustrate the general features by transforming to the AD frame: 
\eqn{
\hat{H}_{U}&\!=\!\left(E_{C}\!-\!\sum_{m=1}^{M-1}\frac{\zeta_{m}^{2}}{\Omega_{m}}\right)\hat{N}^{2}\!+\!V(\varphi+\hat{\Xi})\!+\!\sum_{m}\Omega_{m}\hat{b}_{m}^{\dagger}\hat{b}_{m}\nonumber\\&=\begin{cases}
:V(\hat{\Xi}):\!+\!\sum_{m}\Omega_{m}\hat{b}_{m}^{\dagger}\hat{b}_{m}+O\left(\frac{1}{M}\right) & \eta\leq1\\
E_{C}\frac{\eta-1}{\eta-\frac{1}{2}}\hat{N}^{2}\!+\!V(\varphi+\hat{\Xi})+\sum_{m}\Omega_{m}\hat{b}_{m}^{\dagger}\hat{b}_{m} & \eta>1
\end{cases},\nonumber\\
}
where we define the dimensionless coupling strength (which basically corresponds to $g/\omega$ in the previous notation) by
\eqn{
\eta\equiv\frac{\alpha E_{C}}{\pi W},
}
and use 
$\hat{H}_U\!=\!\hat{U}^\dagger\hat{H}_{\rm C}\hat{U}$ with $\hat{U}\!=\!\exp(-i\hat{N}\hat{\Xi})$ and $\hat{\Xi}\!=\!\sum_mi(\hat{b}^\dagger_m\!-\!\hat{b}_m)\zeta_m/\Omega_m$. Importantly, the effective ``mass" in the transformed frame shows the infrared divergence in $\eta\leq1$, while it remains finite in the ESC regime $\eta>1$. Thus, following our arguments above, we expect that as the coupling strength is increased there occurs the delocalization-localization transition in $\eta\leq1$, while the ground state should ultimately exhibit the reentrant transition to the delocalized phase above $\eta\!\sim\!1$.

To make these predictions concrete, in Fig.~\ref{fig_frg} we show the ground-state phase diagram of the circuit QED system~\eqref{Hcir}, which is obtained by the FRG analysis with the local potential approximation (see, e.g., Refs.~\cite{DUPUIS20211,KM21,TY22} for technical details). In the limit of deep potential barrier $1/v\to 0$, $\varphi_0$ always remains to be positive during the RG flows and thus the localized phase is realized. As the potential barrier becomes shallow ($1/v$ increases), the value of $\varphi_0$ is eventually renormalized to zero in the IR limit and the transition to the delocalized phase occurs. Notably,  the behavior of the transition point drastically changes around $\eta=1$ (vertical dashed line); in $\eta<1$, the localized phase expands as the coupling strength $\eta$ is increased, while the ordering is suppressed in the ESC regime $\eta>1$. This is consistent with our arguments based on the AD frame above and also with the nonmonotonic dependence of  $\zeta_m$ on the coupling strength $\alpha$ in Eq.~\eqref{zetacp}. We expect that these results might be tested by recent experiments realizing galvanic coupling of Josephson junctions to a high-impedance long transmission line \cite{KR19,KR21}.

\section{Summary and Discussions\label{sec:sum}}
We analyzed equilibrium and dynamical properties of light-matter systems consisting of quantum emitters strongly interacting with quantized electromagnetic continuum in the nonperturbative regimes, including the previously unexplored deep and extremely strong coupling regimes. There, traditional theoretical approaches utilizing the Coulomb or PZW gauges are no longer sufficient, since substantial light-matter entanglement invalidates truncations of emitter/photon levels in these gauges. We resolved this problem by using the unitary transformation~\eqref{unitary} that asymptotically disentangles emitter and photon degrees of freedom in the strong-coupling limit; this new frame of reference then enabled us to construct an accurate theoretical framework at any finite interaction strengths. Below we summarize our key findings. 

We first analyzed the single-emitter system (see Eq.~\eqref{HU1}), and elucidated the essential features in the nonperturbative regimes on the basis of general arguments. In particular, we demonstrated the emergence of a ladder of many-body bound states and the (quasi) BIC, the vacuum fluctuations induced suppression of potential barrier, and the strong renormalization of the effective mass. We then analyzed these nonperturbative features in a concrete model of cavity-array waveguide. All of these results  have relevance to ongoing experiments in superconducting qubits interacting with microwave resonators or atoms coupled to photonic crystals. 
We proposed that the BIC can experimentally be observed by analyzing nonequilibrium dynamics induced by the quench of a parameter in the waveguide QED Hamiltonian. This protocol should be implemented in circuit QED systems using currently available experimental techniques.
The parameter regimes we studied are either directly relevant to state-of-the-art experimental systems in, e.g., superconducting devices \cite{YF172,FDP17} and plasmonic crystals \cite{MNS20} or (at least) expected to be accessible in the near future in view of rapid developments in achieving stronger light-matter coupling regimes \cite{FP19,SAS21}. 
To explore those nonperturbative regimes in setups of atoms coupled to photonic crystals, one can utilize Rydberg atoms \cite{AGS84,BP14,LQY18} and/or the collective $\sqrt{N}$ enhancement of the light-matter coupling by assembling a large number of weakly coupled components \cite{RD17} as discussed in Sec.~\ref{subsec:loc}. We envision that the predicted tightly localized bound states in the waveguides can be used as a photon storage in quantum information applications.

We next extended the analysis to the multi-emitter case and established a general framework for studying multi-emitter QED systems without relying on uncontrolled approximations or assumptions. Building on this formalism, we argued that the ground-state physics can be understood from the perspective of the transverse-field Ising model~\eqref{ising} but with suitably renormalized parameters. Finally, we analyzed the case of gapless photonic dispersions and showed that a quantum phase transition can in general occur if the renormalized mass diverges in the thermodynamic limit, while transition is not expected when the mass remains finite. There, we found that in certain cases the diamagnetic term leads to the suppression of the symmetry-broken phase, which is reminiscent of no-go theorems for superradiant transition in cavity QED (see e.g., Refs.~\cite{AGM20,GD20}). One surprising consequence of this analysis is the appearance of a reenterant transition into the symmetry-unbroken phase at sufficiently strong coupling, which was absent in the simplified descriptions such as the spin-boson model. These results are also confirmed by the FRG analysis of the circuit QED system.

It is interesting to analyze the ground-state properties of multi-dipole waveguide QED systems in further detail. In particular, the full understanding of a possible superradiant-type transition in the present multi-dipole systems remains as an intriguing open question. At a qualitative level, our analysis appears to indicate that the tendency to superradiance (i.e., localized phase) will be the strongest at intermediate coupling strengths. For instance, as far as the collective mode plays a dominant role and relative motion can be neglected, we expect that the analysis in Sec.~\ref{sec:gless} can be extended to the multi-dipole cases via replacing the effective mass $m_{\rm eff}$ by the collective one $M_{\rm eff}$ (cf. Eq.~\eqref{collechammat}). If the bosonic dispersion is gapless and $M_{\rm eff}$ exhibits the infrared divergence, then the ground state may exhibit the ordering akin to the superradiant phase, in a close analogy with the delocalization-localization transition for the single-dipole cases discussed in this paper. One of the conceptual advantages in our approach in this respect is to connect these seemingly unrelated phenomena.

Taking the limit of many emitters should provide alternative way to realize a quantum phase transition. In particular, one may use the two-level effective model~\eqref{ising} to determine the ground-state properties in such cases. This mapping to the transverse-field Ising model  suggests the possibilities of inducing a transition between the disordered (i.e., delocalized) phase and the ferromagnetic ordered (i.e., localized) phase or realizing exotic phases such as the many-body localized phase.

Several further open questions also remain for future studies. First, it merits further study to figure out how losses and decoherence either for emitters or photons can affect the present results. They essentially broaden the absorption spectra shown in  Fig.~\ref{fig_quench}(g-i) and can affect the dynamics as excitations acquire finite lifetimes. While the waveguide coupling efficiency can be made close to the unit fidelity in, e.g., microwave superconducting devices \cite{MM19}, those loss effects are still ubiquitous in atoms coupled to photons in optical domains \cite{GA15}. These issues can be addressed by combining the present nonperturbative QED formalism with the standard framework of Markovian open systems \cite{YAreview}. 
Second, instead of the exact diagonalization performed here, one can apply more efficient numerical methods, such as the  matrix-product-states calculations \cite{PB13,SBE14,TS18,BK21} or a hybrid variational approach \cite{WY20}, to analyze the asymptotically decoupled QED Hamiltonian within the few-photon ansatz~\eqref{fewphoton}. This should be particularly useful when one is interested in a larger system with many emitters being coupled to common multiple electromagnetic modes. 
Finally, while the emphasis was placed on the waveguide QED in this paper, our theory is equally applicable to cavity QED setups with multiple photonic modes (see e.g., Ref.~\cite{VVD18}) whose inclusion is often important depending on the cavity geometry and the coupling strength. We also note that the present formalism can be extended to higher-dimensional systems (see, e.g., the Supplementary Materials of Ref.~\cite{YA21}). One intriguing direction is to explore a possible extension of these formalisms to the case in which matter degrees of freedom consist of indistinguishable quantum particles.

Our study is also relevant to a variety of strongly coupled light-matter systems recently realized by using both solid state and quantum chemistry platforms. In particular, in the case of localized $N$ identical emitters, our results obtained for a single-emitter system are expected to remain the same (except for the $\sqrt{N}$ enhancement) as far as the relative motion does not play a significant role (see Sec.~\ref{subsec:loc}). In this respect, the predicted vacuum-induced suppression of the potential barrier in Sec.~\ref{subsec:vac} may lie at the heart of the enhanced chemical reactivity observed in polaritonic chemistry \cite{Hiura2018,Hiura2019,TA19}. More generally, our study reveals that the mass enhancement is one of the universal features of strongly interacting light-matter systems. This naturally associates with the higher density of states, which could lead to enhancements of certain many-body properties, including superconductivity or ferromagnetism \cite{TA192,TA212,Basovaag1992,CJB19,SF19,SMA18,MK19,LJ202,CA15}.

\begin{acknowledgments}
We are grateful to Andrea Cavalleri, Ignacio Cirac,  Jerome Faist, Victor Galitski,  Dieter Jaksch, Israel Klich, Kanta Masuki, Masaki Oshikawa, Angel Rubio, Hossein Sadeghpour, and Tao Shi for useful discussions. Y.A. acknowledges support from the Japan Society for the Promotion of Science through Grant No.~JP19K23424. T.Y. acknowledges support from the RIKEN Special Postdoctoral Researchers Program. E.D. acknowledges support from Harvard-MIT CUA, AFOSR-MURI Photonic Quantum Matter (Grant No.~FA95501610323), the ARO grant ``Control of Many-Body States Using Strong Coherent Light-Matter Coupling in Terahertz Cavities", the NSF EAGER-QAC-QSA Grant No.~2038011 ``Quantum Algorithms for Correlated Electron-Phonon System", and the hospitality of the physics department at ETH Zurich.
\end{acknowledgments}

\appendix
\section{Diagonalization of the quadratic photon Hamiltonian\label{app_diag}}
Here we provide details about the diagonalization of the quadratic photon Hamiltonian including the $\hat{A}^2$ term in the Coulomb gauge. 
To this end, we introduce the conjugate pairs of variables via
\eqn{
\hat{X}_{k}&=&\sqrt{\frac{\hbar}{2\omega_{k}}}\left(\hat{a}_{k}+\hat{a}_{k}^{\dagger}\right),\label{appx}\\
\hat{P}_{k}&=&i\sqrt{\frac{\hbar\omega_{k}}{2}}\left(\hat{a}_{k}^{\dagger}-\hat{a}_{k}\right),\label{appp}
}
and rewrite the quadratic part of $\hat{H}_{\rm C}$ in Eq.~\eqref{HC} as
\eqn{
&&\frac{q^{2}\hat{A}^{2}}{2m}+\sum_{k}\hbar\omega_{k}\hat{a}_{k}^{\dagger}\hat{a}_{k}\nonumber\\
&=&\sum_{k}\frac{\hat{P}_{k}^{2}}{2}\!+\!\frac{1}{2}\sum_{kk'}\left(\delta_{kk'}\omega_{k}^{2}\!+\!2g_{k}g_{k'}\right)\hat{X}_{k}\hat{X}_{k'}.\label{appeqa2}
}
The last term in Eq.~\eqref{appeqa2} can readily be diagonalized by an orthogonal transformation,
\begin{equation}
\hat{P}_{k}=\sum_{n}O_{kn}\hat{\tilde{P}}_{n},\;\;\hat{X}_{k}=\sum_{n}O_{kn}\hat{\tilde{X}}_{n},
\end{equation}
where an orthogonal matrix $O$ satisfies Eq.~\eqref{orthogonal}.
The quadratic photon Hamiltonian then becomes
\begin{equation}
\frac{1}{2}\sum_{n}\left(\hat{\tilde{P}}_{n}^{2}+\Omega_{n}^{2}\hat{\tilde{X}}_{n}^{2}\right)=\sum_{n}\hbar\Omega_{n}\hat{b}_{n}^{\dagger}\hat{b}_{n},
\end{equation}
where we introduce the annihilation operator after the orthogonal transformation by
\begin{equation}
\hat{b}_{n}=\sqrt{\frac{\Omega_{n}}{2\hbar}}\hat{\tilde{X}}_{n}+\frac{i}{\sqrt{2\hbar\Omega_{n}}}\hat{\tilde{P}}_{n},\label{appbop}
\end{equation}
which gives Eq.~\eqref{sqz}. Using these squeezed photon operators  $\hat{b}_n$, the total Hamiltonian in the Coulomb gauge is now given by Eq.~\eqref{HCb} in the main text.

\section{Numerical diagonalization \label{app:num}}
\begin{figure*}[t]
\includegraphics[width=180mm]{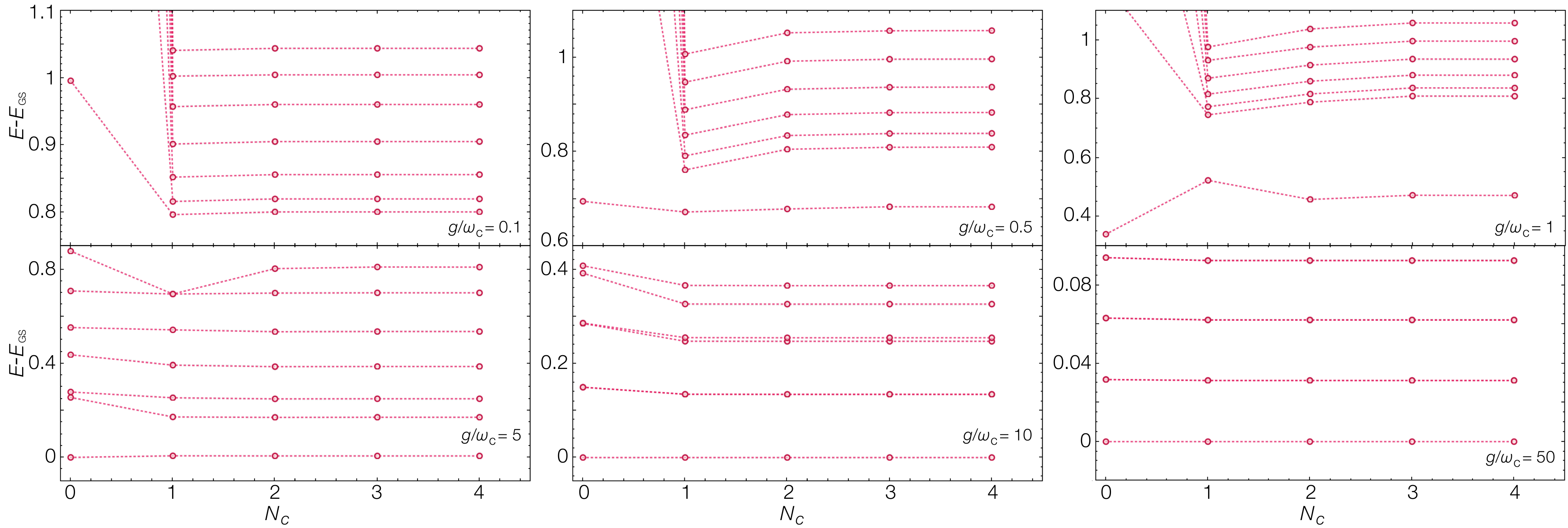} 
\caption{\label{fig_nc}
Convergence of the excitation energies against the total photon-number cutoff $N_c$ at different coupling strengths $g$. The results are obtained by the exact diagonalization of the transformed Hamiltonian~\eqref{ADham} within the few-photon ansatz~\eqref{fewphoton} with the total photon-number cutoff $N_c$. Parameters are  $J=0.2$, $v=0.5$, and $d=0.87$.
}
\end{figure*}
We describe details about the method used in Sec.~\ref{sec:app} to numerically diagonalize the QED Hamiltonian in the AD frame. 
We begin with folding the electromagnetic modes of the cavity array onto even and odd modes with respect to the spatial parity. Since only the even modes,
\begin{equation}
\hat{a}_{x=0}^{e}=\hat{a}_{x=0},\;\;a_{x>0}^{e}=\frac{1}{\sqrt{2}}(\hat{a}_{x}+\hat{a}_{-x}),
\end{equation}
interact with the emitter, we neglect contributions from the odd modes.
The photon Hamiltonian in the real-space basis is then written as
\eqn{
\hat{H}_{{\rm light}} && =-\frac{J}{\sqrt{2}}\left(\hat{a}_{x=0}^{\dagger e}\hat{a}_{x=1}^{e}+{\rm H.c.}\right)\nonumber\\
&&-\frac{J}{2}\sum_{x=1}^{(L-1)/2}\left(\hat{a}_{x+1}^{\dagger e}\hat{a}_{x}^{e}+{\rm H.c.}\right)\!+\!\hbar\omega_{c}\sum_{x=0}^{(L-1)/2}\hat{a}_{x}^{\dagger e}\hat{a}_{x}^{e} \label{appeqd}\nonumber\\
&&\equiv\sum_{p=0}^{(L-1)/2}\hbar\omega_{p}\hat{a}_{p}^{\dagger}\hat{a}_{p},
}
where we use an orthogonal matrix $M$ to obtain the diagonalized form~\eqref{appeqd} with
\begin{equation}
\hat{a}_{x}^{e}=\sum_{p=0}^{(L-1)/2}M_{xp}\hat{a}_{p}.
\end{equation}
The vector field is expressed in this basis as
\eqn{
\hat{A}_{x=0}&=&{\cal A}(\hat{a}_{x=0}+\hat{a}_{x=0}^{\dagger})\nonumber\\
&=&\sum_{p=0}^{(L-1)/2}{\cal A}M_{0p}\left(\hat{a}_{p}+\hat{a}_{p}^{\dagger}\right),
}
which results in the electromagnetic amplitudes (cf. Eq.~\eqref{vecpot})
\begin{equation}
f_{p}={\cal A}M_{0p}.
\end{equation}

We use the frequencies $\omega_p$ and the amplitudes $f_p$ to diagonalize the photon Hamiltonian including the $\hat{A}^2$ term as explained in Appendix~\ref{app_diag}. Then, we perform the unitary transformation~\eqref{unitary} and arrive at the Hamiltonian in the AD frame 
\eqn{
\hat{H}_{U}&=&\frac{\hat{P}^{2}}{2m_{{\rm eff}}}+V_{{\rm eff}}(\hat{Q})\nonumber\\
&+&\sum_{l=1}\frac{:\hat{\Xi}^{l}:}{l!}V^{(l)}(\hat{Q})+\!\sum_{n=0}^{(L-1)/2}\hbar\Omega_{n}\hat{b}_{n}^{\dagger}\hat{b}_{n}.\label{ADham}
}
To numerically diagonalize this Hamiltonian efficiently, we employ the few-photon ansatz~\eqref{fewphoton} and express the matrix elements of $\hat{H}_U$ in terms of the basis
\begin{equation}
|\psi_{\alpha}\rangle_{\rm emitter}\otimes|n_{0}n_{1}\cdots n_{(L-1)/2}\rangle_{\rm photon}
\end{equation}
with level truncations, 
\begin{equation}
\alpha=1,2,\ldots,\alpha_{c},\;\;\sum_{j=0}^{(L-1)/2}n_{j}\leq N_{c}.
\end{equation}
We recall that $|\psi_\alpha\rangle$ are single-particle eigenstates of the renormalized emitter Hamiltonian~\eqref{Hmat}, and $|n_{0}n_{1}\cdots n_{(L-1)/2}\rangle$ is a many-body bosonic Fock state in terms of $\hat{b}_n$ operators with $n_j=0,1,\cdots$. The corresponding Hilbert-space dimension is
\begin{equation}
D=\alpha_{c}\sum_{i=1}^{N_{c}}[(L+1)/2]^{i}=\frac{\alpha_{c}([(L+1)/2]^{N_{c}+1}-1)}{(L+1)/2-1},
\end{equation}
 which grows polynomially with the system size $L$. Figure~\ref{fig_nc} demonstrates that the numerical results converge very efficiently with $N_c$ in a broad range of the light-matter coupling strength. It typically suffices to set $N_c=2$-$4$ and $\alpha_c={\rm O}(10)$ to achieve the accuracy with an error below $\sim \!1\%$. When only the low-energy spectrum is of interest, one can  use the Lanczos method to further reduce the computational cost.

 \section{Derivation of the multi-emitter Hamiltonian in the asymptotically decoupled frame\label{app:multi}}
We here derive the asymptotically decoupled multi-emitter Hamiltonian discussed in Sec.~\ref{sec:cas}.
As we have done for the single-emitter case, we first diagonalize the quadratic photon part of the Coulomb-gauge Hamiltonian including the $\hat{A}^2$ term. To do so,  we introduce the position-dependent multi-emitter coupling strengths by
\eqn{
g_{kj}^{c}&=&qf_{kj}\sqrt{\frac{\omega_{k}}{m_{j}\hbar}}\cos(kx_{j}),\\
g_{kj}^{s}&=&qf_{kj}\sqrt{\frac{\omega_{k}}{m_{j}\hbar}}\sin(kx_{j}),
}
and rewrite the quadratic part as (aside constant)
\eqn{
&&\sum_{j}\frac{q^{2}\hat{A}_{x_{j}}^{2}}{2m_{j}}+\sum_{k}\hbar\omega_{k}\hat{a}_{k}^{\dagger}\hat{a}_{k} \nonumber\\
&=&\sum_{kk'}\left(\delta_{kk'}+2\sum_{j}\frac{g_{kj}^{s}g_{k'j}^{s}}{\omega_{k}\omega_{k'}}\right)\frac{\hat{P}_{k}\hat{P}_{k'}}{2}\nonumber\\
&&+\sum_{kk'}\left(\delta_{kk'}\omega_{k}^{2}+2\sum_{j}g_{kj}^{c}g_{k'j}^{c}\right)\frac{\hat{X}_{k}\hat{X}_{k'}}{2}\nonumber\\
 & &-\sum_{kk'}\sum_{j}\frac{g_{kj}^{c}g_{k'j}^{s}}{\omega_{k'}}\left(\hat{X}_{k}\hat{P}_{k'}+\hat{P}_{k'}\hat{X}_{k}\right),\label{apptemp}
}
where we recall that the conjugate operators $\hat{X}_k$ and $\hat{P}_k$ are defined by Eqs.~\eqref{appx} and \eqref{appp}. 
Equation~\eqref{apptemp} can then be diagonalized by the symplectic transformation,
\begin{equation}
\left(\begin{array}{c}
\hat{\boldsymbol{X}}\\
\hat{\boldsymbol{P}}
\end{array}\right)=S\left(\begin{array}{c}
\hat{\tilde{\boldsymbol{X}}}\\
\hat{\tilde{\boldsymbol{P}}}
\end{array}\right)\equiv\left(\begin{array}{cc}
S^{XX} & S^{XP}\\
S^{PX} & S^{PP}
\end{array}\right)\left(\begin{array}{c}
\hat{\tilde{\boldsymbol{X}}}\\
\hat{\tilde{\boldsymbol{P}}}
\end{array}\right),
\end{equation}
where the matrix $S$ satisfies 
\eqn{
S\sigma S^{\rm T}=\sigma
} 
with $\sigma=i\sigma^y\otimes {\rm I}_L$ and ${\rm I}_L$ being the $L\times L$ identity matrix. This  leads to the diagonalized form
\eqn{\label{appzeta2}
\sum_{j}\frac{q^{2}\hat{A}_{x_{j}}^{2}}{2m_{j}}+\sum_{k}\hbar\omega_{k}\hat{a}_{k}^{\dagger}\hat{a}_{k}&=&\frac{1}{2}\sum_{n}\left(\hat{\tilde{P}}_{n}^{2}+\Omega_{n}^{2}\hat{\tilde{X}}_{n}^{2}\right)\nonumber\\
&=&\sum_{n}\hbar\Omega_{n}\hat{b}_{n}^{\dagger}\hat{b}_{n},
}
where we define the squeezed photon operators $\hat{b}_n$ in the same manner as in Eq.~\eqref{appbop}.

In terms of these new photon operators, the Coulomb-gauge Hamiltonian is expressed as
\eqn{\label{apphc}
\hat{H}_{{\rm C}}&=&\sum_{j}\left[\frac{\hat{P}_{j}^{2}}{2m_{j}}+V(\hat{Q}_{j})\right]\nonumber\\
&&-\sum_{jn}\hat{P}_{j}\left(\zeta_{nj}^{*}\hat{b}_{n}+\zeta_{nj}\hat{b}_{n}^{\dagger}\right)+\sum_{n}\hbar\Omega_{n}\hat{b}_{n}^{\dagger}\hat{b}_{n},
}
where we define
\eqn{\label{appzeta}
\zeta_{nj}&=&\sqrt{\frac{\hbar}{m_{j}\Omega_{n}}}\nonumber\\
&\times&\sum_{k}\left[g_{kj}^{c}S_{kn}^{XX}-g_{kj}^{s}S_{kn}^{PX}+i\left(g_{kj}^{c}S_{kn}^{XP}-g_{kj}^{s}S_{kn}^{PP}\right)\right].\nonumber\\
}

We now introduce the multi-emitter extension of the asymptotically decoupling unitary transformation by
\eqn{
\hat{U}&=&\exp\Biggl[\frac{1}{\hbar}\sum_{jn}\hat{P}_{j}\frac{(\zeta_{nj}\hat{b}_{n}^{\dagger}-\zeta_{nj}^{*}\hat{b}_{n})}{\Omega_{n}}\Biggr]\nonumber\\
&\equiv&\exp\Bigl(-\frac{i}{\hbar}\sum_{j}\hat{P}_{j}\hat{\Xi}_{j}\Bigr),
}
where we recall $\hat{\Xi}_{j}=\sum_{n}i(\xi_{nj}\hat{b}_{n}^{\dagger}-\xi_{nj}^{*}\hat{b}_{n})$ with the displacement parameters
\eqn{
\xi_{nj}=\frac{\zeta_{nj}}{\Omega_{n}}.
}
This transformation acts on the photon and emitter operators as
\eqn{
\hat{U}^{\dagger}\hat{b}_{n}\hat{U}&=&\hat{b}_{n}+\sum_{j}\frac{\xi_{nj}\hat{P}_{j}}{\hbar},\\
\hat{U}^{\dagger}\hat{Q}_{j}\hat{U}&=&\hat{Q}_{j}+\hat{\Xi}_{j},
}
and transforms Eq.~\eqref{apphc} to the following form:
\eqn{
\hat{H}_{U}&=&\hat{U}^{\dagger}\hat{H}_{{\rm C}}\hat{U}\nonumber\\
&=&\sum_{j}\left[\frac{\hat{P}_{j}^{2}}{2m_{{\rm eff},j}}+V(\hat{Q}_{j}+\hat{\Xi}_{j})\right]-\sum_{i> j}\mu_{ij}\hat{P}_{i}\hat{P}_{j}\nonumber\\
&&+\sum_{n}\hbar\Omega_{n}\hat{b}_{n}^{\dagger}\hat{b}_{n},
}
which gives Eq.~\eqref{multihu} in the main text. Here, we introduce the effective mass $m_{{\rm eff},j}$ for each emitter and the emitter-emitter coupling $\mu_{ij}$ as
\eqn{
m_{{\rm eff},j}&\equiv&\frac{m_{j}}{1-\sum_{n}\frac{2m_{j}|\zeta_{nj}|^{2}}{\hbar\Omega_{n}}},\\
\mu_{ij}&\equiv&\sum_{n}\frac{2\zeta_{ni}^{*}\zeta_{nj}}{\hbar\Omega_{n}}.
}
The expressions~\eqref{multimeff} and \eqref{multimu} in Sec.~\ref{sec:cas} follow from the relations~\eqref{appzeta2} and \eqref{appzeta}.

\bibliography{reference}

\end{document}